\documentclass[12pt,preprint]{emulateapj}%{aastex}

\usepackage{graphics,graphicx}
\usepackage{subfigure}
\usepackage{color}
\usepackage{placeins}

\begin{document}

\title{On $R-W1$ as A Diagnostic to Discover Obscured Active Galactic Nuclei in Wide-Area X-ray Surveys}

\author{Stephanie M. LaMassa$^{1,2}$, 
Francesca Civano$^{1,3}$,
Marcella Brusa$^{4,5}$,
Daniel Stern$^{6}$,
Eilat Glikman$^{7}$,
Sarah Gallagher$^{8}$,
C. Meg Urry$^{1,2}$,
Sabrina Cales$^{1,2,9}$,
Nico Cappelluti$^{5}$,
Carolin Cardamone$^{10}$,
Andrea Comastri$^{5}$,
Duncan Farrah$^{11}$,
Jenny E. Greene$^{12}$,
S. Komossa$^{13}$,
Andrea Merloni$^{14}$,
Tony Mroczkowski$^{15,16}$,
Priyamvada Natarajan$^{1,2,17}$,
Gordon Richards$^{18}$,
Mara Salvato$^{14}$,
Kevin Schawinski$^{19}$,
Ezequiel Treister$^{9}$}

\affil{$^1$Yale Center for Astronomy \& Astrophysics, Physics Department, P.O. Box 208120, New Haven, CT 06520, USA,
$^2$Department of Physics, Yale University, P.O. Box 208121, New Haven, CT 06520, USA ,
$^3$Smithsonian Astrophysical Observatory, 60 Garden Street, Cambridge, MA 02138, USA,
$^4$DIFA—Dipartimento di Fisica e Astronomia, Universita' di Bologna, viale Berti Pichat 6/2, I-40127 Bologna, Italy, 
$^5$INAF-Osservatorio Astronomico di Bologna, via Ranzani 1, I-40127 Bologna, Italy, 
$^6$Jet Propulsion Laboratory, California Institute of Technology, Pasadena, CA 91109, USA,
$^7$Department of Physics, Middlebury College, Middlebury, VT 05753, USA
$^8$Department of Physics \& Astronomy, The University of Western Ontario, London, ON N6A 3K7, Canada,
$^9$Department of Astronomy, University of Concepcion, Concepcion, Chile,
$^{10}$Department of Math \& Science, Wheelock College, 200 Riverway, Boston, Massachusetts 02215,
$^{11}$Department of Physics MC 0435, Virginia Polytechnic Institute and State University, 850 West Campus Drive, Blacksburg, VA 24061, USA,
$^{12}$Department of Astrophysical Sciences, Princeton University, Princeton, NJ 08544, USA,
$^{13}$Max-Planck-Institut f\"ur Radioastronomie, Auf dem H\"ugel 69, D-53121 Bonn, Germany,
$^{14}$Max-Planck-Institut f\"ur extraterrestrische Physik, Giessenbachstrasse 1, D-85748 Garching bei M\"unchen, Germany, 
$^{15}$Naval Research Laboratory, 4555 Overlook Avenue SW, Washington, DC, USA ,
$^{16}$National Research Council Fellow,
$^{17}$Department of Astronomy, Yale University, P.O. Box 208101, New Haven, CT 06520, USA,
$^{18}$Department of Physics, Drexel University, 3141 Chestnut Street, Philadelphia, PA 19104, USA,
$^{19}$Institute for Astronomy, Department of Physics, ETH Zurich, Wolfgang-Pauli Strasse 27, CH-8093 Zurich, Switzerland
}

\begin{abstract}
Capitalizing on the all-sky coverage of {\it WISE}, and the 35\% and 50\% sky coverage from SDSS and Pan-STARRS, respectively, we explore the efficacy of $m_{R}$ (optical) - $m_{3.4 \mu m}$ (mid-infrared), hereafter $R-W1$, as a color diagnostic to identify obscured supermassive black hole accretion in wide-area X-ray surveys. We use the $\sim$16.5 deg$^2$ Stripe 82 X-ray survey data as a test-bed to compare $R-W1$ with $R-K$, an oft-used obscured AGN selection criterion, and examine where different classes of objects lie in this parameter space. Most stars follow a well-defined path in $R-K$ vs. $R-W1$ space. We demonstrate that optically normal galaxies hosting X-ray AGN at redshifts $0.5<z<1$ can be recovered with an $R-W1>4$ color-cut, while they typically are not selected as AGN based on their $W1-W2$ colors. Additionally, different observed X-ray luminosity bins favor different regions in $R-W1$ parameter space: moderate luminosity AGN ($10^{43}$ erg s$^{-1} < L_{\rm 0.5-10 keV} < 10^{44}$ erg s$^{-1}$) tend to have red colors while the highest luminosity AGN ($L_{\rm 0.5-10 keV} > 10^{45}$ erg s$^{-1}$) have bluer colors; higher spectroscopic completeness of the Stripe 82X sample is needed to determine whether this is a selection effect or an intrinsic property. Finally, we parameterize X-ray obscuration of Stripe 82X AGN by calculating their hardness ratios (HRs) and find no clear trends between HR and optical reddening. Our results will help inform best-effort practices in following-up obscured AGN candidates in current and future wide-area, shallow X-ray surveys, including the all-sky {\it eROSITA} mission.

\end{abstract}

\section{Introduction}
Supermassive black holes growing via accretion (active galactic nuclei, or AGN) are often identified by tell-tale signatures in their spectra, such as Doppler broadened emission lines from gas orbiting near the black hole or strong narrow emission lines in gas hundreds of parsecs away from the black hole, yet still primarily ionized by the AGN continuum. The former ``broad-lined'' AGN are easily discovered by large ground-based optical surveys, such as the Sloan Digital Sky Survey (SDSS), which targeted objects based on optical photometry for follow-up spectroscopy, revealing hundreds of thousands of AGN with ``blue'' colors and point-like morphology \citep{york, dr1, dr10}. Conversely, obscured AGN represent those sources where the active nucleus is enshrouded, and can come in two flavors: objects lacking broad emission lines but displaying powerful narrow emission lines indicative of supermassive black hole accretion, i.e., the Type 2 AGN explained by the AGN unification model \citep{antonucci, urry}, and those sources with broad emission lines but ``red'' colors due to large amounts of dust attenuating and reddening the optical continuum light \citep[e.g.,][]{glikman1}; here the SED at optical wavelengths is suppressed with respect to the near- to mid-infrared emission.

Many accreting supermassive black holes generate powerful X-ray emission visible across the Universe, making X-ray selection an efficient mechanism for obtaining clean AGN samples with virtually no contamination from star-forming galaxies at X-ray luminosities exceeding 10$^{42}$ erg s$^{-1}$ \citep[see][for a recent review]{brandt2015}, complementing the AGN census revealed from optical and infrared surveys. However, as hard X-rays can select both the unobscured and obscured AGN populations (though the most heavily obscured, i.e., Compton-thick, AGN can be missed in X-ray surveys), supporting multi-wavelength data are necessary to identify which objects are the obscured systems missing from our current census of supermassive black hole growth. While optical/infrared spectroscopy is the ``gold standard'' in determining whether or not an X-ray source is an obscured AGN, most X-ray sources lack immediate spectroscopic identifications, and following-up all X-ray objects with spectroscopy is expensive and time-consuming. Establishing widely applicable diagnostics to uncover obscured black hole growth are then important for mining these large X-ray and multiwavelength datasets for the most promising candidates to follow-up, especially as resources are limited. In addition to current and planned wide-area X-ray surveys, such as the 16.5 deg$^2$ Stripe 82X \citep{me1,me2}, the 50 deg$^2$ {\it XMM}-XXL (PI: Pierre),  and the $\sim$877 deg$^2$ {\it XMM}-Serendipitous \citep{xmm_seren5} surveys, {\it eROSITA} will map the entire sky from 0.5-10 keV starting in 2017 \citep{erosita}, revealing millions of AGN candidates \citep{merloni}. 

\subsection{Identifying Type 2 AGN}
A combination of optical and infrared clues have been used to identify Type 2 AGN. Below $z=0.5$, the traditional BPT diagnostic ratios \citep{bpt,vo}, [NII]6584\AA/H$\alpha$ versus [OIII]5007\AA/H$\beta$, trace the ionization potential and are thus effective at separating narrow-lined AGN from star-forming galaxies; tens of thousands of such obscured AGN have been classified by SDSS with this method \citep[e.g.,][]{kauffmann,kewley06}. Recently, He II 4686\AA/H$\beta$ versus [NII] 6584/H$\alpha$ was introduced as an empirical diagnostic to separate star-forming galaxies from AGN \citep{shirazi}. 

As H$\alpha$ is redshifted out of the optical bandpass at $z>0.5$, alternative obscured AGN selection methods using observed optical spectra have been developed to capitalize on the existing rich datasets from wide-area optical surveys, including: 
\begin{itemize}
\item rest-frame $g-z$ color versus [NeIII]/[OII] which can be applied to $z\sim$1.4 \citep[TBT,][]{tbt}; 
\item stellar mass versus [OIII]/H$\beta$, which can select AGN up to $z\sim$1 \citep[Mass-Excitation diagnostic, MEx,][]{mex}; 
\item {[OII]3726+2739\AA/H$\beta$} versus [OIII]/H$\beta$, which is applicable to $z\sim1$ \citep{lamareille}; 
\item strong [NeV]3426\AA\  emission, which can select AGN to $z\sim1.5$ \citep{gilli,mignoli}\footnote{While [NeV] is a doublet with emission at 3426\AA\  and 3326\AA, [NeV]3326\AA\  has a third the intensity of [NeV]3426\AA\ \citep{vandenberk} and has thus not been used to select samples of obscured AGN.}; 
\item a combination of narrow emission lines with full width half-maximum (FWHM) values $<$ 2000 km s$^{-1}$, ratios of narrow emission lines suggesting AGN rather than star-formation ionization, and a high [OIII] luminosity, which is applicable between $0.3<z<0.83$ \citep{zakamska, reyes}; 
\item strong CIV and Ly$\alpha$ emission with FWHM $<$2000 km s$^{-1}$ and weak rest-frame ultraviolet continuum values, which identifies AGN between $2 < z < 4.3$ \citep[][see also \citet{steidel,hainline}]{alexandroff}.
\end{itemize}
 Alternatively, candidate narrow-lined AGN at $z\gtrsim 0.5-1$ can be identified based on signatures at other wavelengths (e.g., infrared color, infrared/optical colors, X-ray emission) and followed-up with ground-based infrared spectroscopy to observe the traditional rest-frame BPT emission lines, where the cosmic BPT diagram can then differentiate between AGN and star-forming galaxies \citep{kewley1,kewley2,kartaltepe}.

\subsection{Selecting Reddened Broad-lined AGN}
Unlike Type 2 AGN, so-called ``red quasars'' have broad emission lines \citep[e.g.,][]{glikman1}, but their optical spectra are attenuated by dust and thus are not generally targeted by optical spectroscopic surveys following-up quasar candidates. These broad lines are generally seen at longer wavelengths (i.e., H$\alpha$ and possibly H$\beta$), which suffers less from dust extinction and host galaxy dilution, and may be reddened Type 1.8 and 1.9 AGN,\footnote{Type 1.8 AGN have weak broad components to H$\alpha$ and H$\beta$ while Type 1.9 AGN have a broad component to the H$\alpha$ line only.} rather than pure Type 1 AGN, since higher-ionization lines such as MgII are narrow or absent \citep[e.g.,][]{brusa15}. These reddened AGN are an interesting class of objects, and at least some appear to be in a transition stage between an obscured and unobscured phase within the merger-induced galaxy/black hole co-evolution paradigm \citep{sanders,hopkins,brusa2010,glikman3,banerji1,farrah}. Indeed, {\it Hubble} followed-up a handful of these objects and revealed that they live in train-wreck host galaxies, indicative of merger activity \citep{urrutia,glikman15}. Recent VLT X-shooter observations of a sample of 10 red quasars unearthed powerful outflows in most of these systems, attributed to AGN winds that may be clearing out obscuring dust \citep{brusa15}. Unlike Type 2 AGN selected via BPT diagnostics, that are generally lower-luminosity objects, such red quasars are the highest luminosity AGN at every redshift \citep{glikman3,banerji2}.

Combining optical and infrared data is a powerful tool in the search for obscured AGN, including Type 2 sources and reddened broad-lined objects: dust that attenuates optical signatures of black hole growth re-radiates this emission in the near- to mid-infrared ($\sim1-60\mu$m), producing visible effects on the SED. Thus an object that is optically faint while being infrared bright is a candidate hidden black hole. For instance, an oft-used diagnostic to identify red quasars is a color of $R-K>4-5$ \citep{glikman1}; this has been quite successful in identifying samples of reddened AGN that are missed by optical-only surveys \citep[e.g.,][]{brusa1,brusa15,glikman2,glikman3,glikman4,georgakakis,banerji1,banerji2}. 

Obscured AGN, both narrow-lined and reddened broad-lined objects, have also been identified using:
\begin{itemize}
\item $R-[3.6]$ in tandem with log($\nu_{\rm 24}F_{\rm 24}/\nu_{R}F_{R}$), where [3.6] is the {\it Spitzer} magnitude at 3.6$\mu$m \citep{fiore2008,fiore2009,yan,melbourne}; 
\item $R-[4.5]$, where [4.5] is the {\it Spitzer} magnitude at 4.5$\mu$m \citep{hickox}; 
\item $R-W2$, where $W2$ is the {\it Wide-Field Infrared Survey Explorer} \citep[{\it WISE};][]{wise} flux at 4.6 $\mu$m \citep{stern12,yan2013,donoso}.
\end{itemize} 
More recently, $R-W4$, where $W4$ is the {\it WISE} flux at 22$\mu$m, used in conjunction with non-detections in the {\it WISE} $W1$ (3.4 $\mu$m) and $W2$ bands \citep[see][for a discussion of $W1-W2$ dropout sources]{eisenhardt,wu,stern}, have been shown to identify extremely reddened Type 1 quasars \citep{yan2013, ross_2014}, though such objects are rare and have a low space density. Additionally, infrared-only colors, such as {\it Spitzer} IRAC color selection \citep{lacy2004,stern2005,donley} and WISE $W1-W2$ color selection \citep{stern12,assef}, have been adopted to identify both unobscured and obscured AGN, including Type 2 AGN locally and at high-redshift and high-luminosity \citep[e.g.,][]{lacy2015}, though these color selections work best at brighter fluxes since star-forming galaxy contamination becomes significant at fainter fluxes \citep{barmby,donley2008,cardamone,mateos,mendez}. 

\subsection{Widening the Multi-wavelength Search to Larger Areas}
Though previous optical-mid-infrared colors have been successful in illuminating a unique population of obscured AGN, they can be limited in scope. For instance, {\it Spitzer} observations cover a relatively small area of the sky, limiting the $R-[3.6]$ and $R-[4.5]$ diagnostics to existing {\it Spitzer} survey areas, such as the {\it Spitzer} Wide-area InfraRed Extragalactic Survey\citep[SWIRE;][]{lonsdale}, the Great Observatories Origins Deep Survey \citep[GOODS;][]{goods}, COSMOS \citep{sanders_cosmos}, the {\it Spitzer} Deep, Wide-Field Survey \citep[SDWFS;][]{ashby}, the {\it Spitzer} South Pole Telescope Deep Field \citep[SSDF;][]{ssdf}, the {\it Spitzer} Extragalactic Representative Volume Survey \citep[SERVS;][]{servs}, and the {\it Spitzer} Public Legacy Survey of the UKIDSS\footnote{UKIDSS is the UKIRT Infrared Deep Sky Survey\citep{ukidss}.} Ultra Deep Survey (SpUDS; PI: J. Dunlop). 

The Two-Micron All-Sky Survey \citep[2MASS;][]{2mass} was an all-sky near-infrared survey, allowing samples of $R-K$-selected candidates to be studied \citep[e.g.,][]{glikman1}, but at a relatively shallow depth compared with UKIDSS. Though the UKIDSS Large Area Survey reaches $\sim$1.5 magnitudes deeper than 2MASS in $K$-band, it is limited in area to $\sim$7500 deg$^2$. The VISTA Hemisphere Survey \citep{mcmahon} is surveying the Southern Hemisphere in $J$ and $K$, reaching flux limits 30 times deeper than 2MASS. However, a significant fraction of the sky observed by {\it eROSITA} will lack this deeper $K$-band coverage.

Taking advantage of the all-sky coverage of {\it WISE}, where the $W1$ is the most sensitive {\it WISE} band, and the 35\% and 50\% sky coverage of SDSS \citep{york} and Pan-STARRs \citep{panstarrs}, respectively, we investigate $R-W1$ as a diagnostic to uncover obscured supermassive black hole growth in wide-area X-ray surveys. While past studies used $R-W2>6$ to identify the most obscured AGN in a sample of $W1-W2$-selected objects \citep{yan2013,donoso}, here we take a complementary approach by defining AGN on the basis of their X-ray, rather than infrared, properties, ensuring no contamination from star-forming galaxies in our AGN sample. 

Our parent sample is drawn from the Stripe 82X survey which has a relatively high level of spectroscopic completeness ($\sim$30\% for all Stripe 82X sources) compared with all-sky surveys where supporting spectroscopic coverage is inhomogenous. By focusing on objects spectroscopically identified in Stripe 82X, using observed X-ray luminosities to differentiate between AGN and galaxies without active black holes, we compare $R-W1$ with $R-K$ to highlight where different classes of objects live in this parameter space. We demonstrate that optically elusive X-ray AGN beyond the local Universe ($z>0.5$) can be recovered on the basis of their red $R-W1$ colors. Additionally, as the Stripe 82X sources are likely to be representative of the objects that are initially discovered in current and future wide-area X-ray surveys prior to follow-up observations, we report on the $R-W1$ colors for different observed X-ray luminosity bins.  We then compare the mid-infrared only $W1-W2$ AGN color selection with our X-ray sample to determine which portion of the AGN population are recovered by this color cut. Finally, we test whether inferred X-ray obscuration is associated with optical reddening to determine whether dust that obscures optical light plays a role in attenuating X-ray emission. Throughout the paper, we report $R-W1$ and $R-K$ colors in the Vega magnitude system and adopt a cosmology where H$_{o}$ = 70 km s$^{-1}$ Mpc$^{-1}$, $\Omega_M$ = 0.27 and $\Lambda$=0.73 throughout the paper. Additionally, all reported X-ray luminosities represent observed-frame, non-absorption corrected, full-band (0.5-10 keV)\footnote{While the {\it XMM-Newton} full band ranges from 0.5-10 keV, the {\it Chandra} full-band coverage is from 0.5-7 keV, as {\it Chandra} is much less sensitive above 7 keV.} luminosities.

\section{Data Analysis}
\subsection{The Stripe 82 X-ray Sample}
Stripe 82X is an ongoing {\it XMM-Newton} and {\it Chandra} survey of the legacy SDSS Stripe 82 field: in addition to being imaged over 80 times in the optical, with coadded photometry 1.9-2.2 times deeper than any single-epoch SDSS region \citep{jiang_coadd}, it has a rich investment of optical spectroscopy from SDSS and SDSS BOSS \citep[Data Releases 9 and 10;][]{sdss_dr9,sdss_dr10}, 2 SLAQ \citep{croom},  WiggleZ \citep{drinkwater}, DEEP2 \citep{newman}, PRIMUS \citep{coil}, the deep spectroscopic survey of faint quasars from \citet{jiang}, and a pre-BOSS pilot survey using Hectospec on MMT \citep{ross_mmt}. Furthermore, it has UKIDSS coverage, allowing for deeper near-infrared coverage compared with the all-sky 2MASS survey. Archival {\it Chandra} and {\it XMM-Newton} observations in the field cover $\sim$13.3 deg$^2$, while an additional 4.6 deg$^2$ was obtained from an AO10 {\it XMM-Newton} program, resulting in a total of $\sim$16.5 deg$^2$ of non-overlapping area \citep[PI: C. M. Urry;][]{me1,me2}; an additional $\sim$15.6 deg$^2$ of Stripe 82X was observed by {\it XMM-Newton} in AO13 \citep[PI: C. M. Urry;][]{me_ao13}. As a wide-area survey, the X-ray depth is largely shallow, reaching a full-band flux limit of 5.6 $\times 10^{-15}$ erg s$^{-1}$ cm$^{-2}$, with a half-area flux limit of 1.6 $\times 10^{-14}$ erg s$^{-1}$ cm$^{-2}$, for the initial 16.5 deg$^2$ Stripe 82X release. 

In total, 3362 X-ray sources make up the Stripe 82X sample, detected at the $\geq$5$\sigma$ level with {\it XMM-Newton} or $\geq$4.5$\sigma$ level with {\it Chandra} \citep{me1,me2}. As described in \citet{me2}, the optical and infrared counterparts to the X-ray sources were found using the maximum likelihood estimator, which is a statistical algorithm which accounts for the distance between an X-ray source and multi-wavelength objects found within the search radius (5$^{\prime\prime}$ for {\it Chandra} and 7$^{\prime\prime}$ for {\it XMM-Newton}), as well as the magnitude distribution of sources in the background and the positional errors on the X-ray and multi-wavelength sources \citep{mle}. 

\subsection{\label{samp} $R-W1$ Sample}
Of the 3362 Stripe 82X sources, we focus on the objects that have spectroscopic redshifts so that we can identify the sources and have optical classifications where possible. About 30\% of the Stripe 82X sample is spectroscopically complete, leaving 1005 objects, of which 146 were obtained from our dedicated follow-up campaign with WIYN HYDRA and Palomar DoubleSpec from 2012 to 2014. Although many sources for which we have optical spectroscopic redshifts from existing surveys were targeted based on their optical properties, we have made no such {\it a priori} cuts on our candidate lists for our follow-up observations so that we achieve as fair a sampling of the X-ray population as possible: we have an on-going spectroscopic campaign to target every X-ray source that has an optical counterpart. Though by limiting our sample to sources that have existing spectroscopic redshifts imposes a bias based on optical properties, the AGN demographics will be representative of those objects immediately identified in current and future wide-area X-ray surveys as these surveys will also be affected by limited spectroscopic completeness.  

We then impose an optical magnitude cut of $r\leq22.2$ (AB) and $i\leq$21.3 (AB), which represents the 95\% completeness limit for point sources in the single-epoch SDSS catalog, garnering 878 objects. From this sample, we then retain objects matched to {\it WISE} that have significant $W1$ detections (SNR $>$ 5) and {\it WISE} coordinates within 3$^{\prime\prime}$ of the SDSS coordinates: as the multi-wavelength counterparts were matched to the X-ray source list independently in \citet{me2}, this additional positional cut mitigates unrelated associations between the SDSS and {\it WISE} sources. Thus, as summarized in Table \ref{xray_srcs}, our $R-W1$ sample contains 661 objects; 76 of these were spectroscopically identified by our follow-up campaign.

\subsubsection{UKIDSS Subsample}
In this work, we compare how $R-W1$ relates to $R-K$, an often used color to identify reddened AGN. We then define a UKIDSS subsample that has $K$-band detections and UKIDSS coordinates within 2$^{\prime\prime}$ of the SDSS coordinates, where a smaller search radius is used here compared with {\it WISE} due to the smaller PSF and more precise astrometry of UKIDSS. As noted in Table \ref{xray_srcs}, this subsample contains 552 objects.

\subsection{Object Classification}
SDSS and other optical surveys provide automatic classifications of sources based on their optical spectra, and we have adopted their methodology when characterizing objects observed with our dedicated follow-up campaigns: Type 1 AGN have at least one broad line (generally, FWHM $>$ 2000 km s$^{-1}$) present in their optical spectrum; galaxies are objects which lack any broad lines, but these include narrow-lined AGN, blazars which have featureless optical spectra,\footnote{Though blazars are AGN, they can be mis-categorized as ``galaxies'' by optical classification pipelines due to their lack of broad lines. Visual inspection of the spectra can correctly identify these sources, which we discuss further in Section \ref{opt_gal}.} as well as objects without any signatures of accretion in their optical spectra; and stars are sources with absorption or emission features at $z\sim0$. Based on the optical spectra, 499 of the 661 X-ray objects are Type 1 AGN, 129 are optical galaxies, 30 are stars, and three do not have optical classifications in the ground-based databases we mined. 

In this work, we refer to X-ray AGN as those sources with observed 0.5-10 keV luminosities exceeding 10$^{42}$ erg s$^{-1}$, or if the full-band flux is not significant, 0.5-2 keV or 2-10 keV luminosities above 10$^{42}$ erg s$^{-1}$.\footnote{Seventeen extragalactic X-ray objects have non-significant full-band fluxes but significant soft (0.5-2 keV) and hard (2-10 keV) band fluxes. We consider these sources X-ray AGN if either their soft- or hard-band luminosities exceed 10$^{42}$ erg s$^{-1}$.} Star-formation processes generally are not energetic enough to produce X-ray emission beyond this limit \citep[e.g.,][]{brandt,brandt2015}, though the extended hot gas halo of passive galaxies can reach higher luminosities at $z>0.55$ \citep[e.g.,][]{civano2014}. Of the 661 objects in our sample, 612 are X-ray AGN and 19 are X-ray galaxies, which are also optically classified as galaxies. We caution, however, that the 10$^{42}$ erg s$^{-1}$ luminosity cut omits low-luminosity and heavily X-ray obscured AGN from our sample.

We summarize the optical and X-ray classifications of our $R-W1$ sample and UKIDSS subsample in Table \ref{r-w1_samp}.

\begin{table}[h]
   \begin{center}
    \caption{\label{xray_srcs}Summary of Stripe 82X Sample}
   \begin{tabular}{lr} 
   \hline
   \hline
    & Number\\
    \hline

   Stripe 82X All & 3362 \\
\hline
 Spectroscopic redshifts                                       & 1005 \\
 $r \leq 22.2$,$i \leq 21.3$ (AB)                           &  878  \\
{\it WISE} counterparts  ($R-W1$ sample)             &  661  \\
$R-W1$ sample with UKIDSS counterparts           &  552  \\
\hline
\hline
   \end{tabular}
   \end{center}
\end{table}

\begin{table}[h]
   \begin{center}
    \caption{\label{r-w1_samp}Summary of $R-W1$ Sample \& UKIDSS Subsample}
\begin{tabular}{lrr}
   \hline
   \hline
    Class & $R-W1$ & UKIDSS \\
    \hline
\multicolumn{3}{c}{{\it X-ray Classification}} \\
X-ray AGN ($L_{\rm x} > 10^{42}$ erg s$^{-1}$)       & 612 & 507 \\
X-ray Galaxies ($L_{\rm x} < 10^{42}$ erg s$^{-1}$) &  19  &   17 \\
\hline
\multicolumn{3}{c}{{\it Optical Classification}} \\
Broad-lined AGN                                           & 499 & 406 \\
Galaxies (no broad-lines in spectrum)   & 129  & 115\\
Stars & 30 & 28 \\
No classification & 3 & 3 \\
\hline
\multicolumn{3}{c}{{\it X-ray \& Optical Classification}} \\
Obscured AGN$^{1}$                                      & 110 & 98 \\
Obscured AGN, $z>0.5$                                &  19  & 17 \\
\hline
\hline
\multicolumn{3}{l}{$^{1}$In this context, obscured AGN have optical spectra classified}\\
\multicolumn{3}{l}{as ``galaxies'' and X-ray luminosities exceeding 10$^{42}$ erg s$^{-1}$.}
   \end{tabular}
   \end{center}
\end{table}

\subsection{\label{sdss_mag} Calculating $R-W1$}
\begin{figure}
\begin{center}
{\includegraphics[scale=0.4]{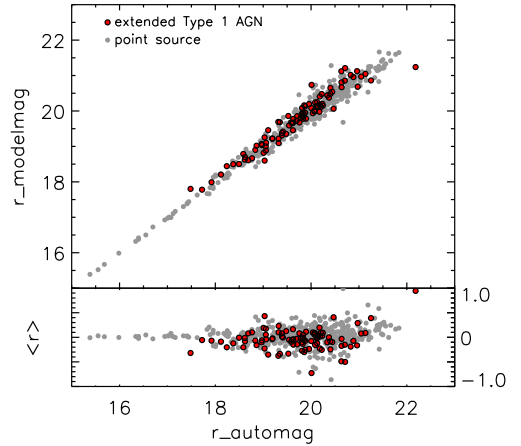}}
\caption[]{\label{comp_mags} {\it Top}: SDSS $r$-band ``modelMag'' magnitude (PSF for point sources; better of a de Vaucouleurs or exponential profile fit for extended sources) compared with the SExtractor ``auto'' magnitude from the \citet{jiang_coadd} coadded catalog and ({\it bottom}) the difference between the ``auto'' and ``modelMag'' magnitudes ($<r>$). The Type 1 AGN with extended photometry are shown by the filled red circles while the point sources are shown by the grey circles. Though the flux from a Type 1 AGN point source can cause the ``modelMag'' magnitude to be a poor representation of total emission when the source is extended, while ``auto'' magnitude is most reliable for such ojects, significant differences between the two measurements are not apparent in our subsample of Type 1 AGN with extended photometry. }
\end{center}
\end{figure}

To develop our diagnostic color selection, we utilize the SDSS ``modelMag'' magnitude. For point sources, this is a PSF model while for extended sources it is the better of a de Vaucouleurs profile fit or an exponential profile fit; $\sim$66\% of the 661 sources in our sample have morphologies consistent with a point source. For the remaining 34\% of sources with extended photometry, the modelMag may be an unreliable estimate of the total magnitude for type 1 AGN, as the AGN point source flux can contribute significantly to the total emission, but be poorly modeled by a de Vaucouleurs or exponential profile fit. To test this effect, we compared the ``modelMag'' with the ``auto'' magnitudes reported in the \citet{jiang_coadd} co-added catalog, which were measured with SExtractor \citep{sextractor}; such ``auto'' magnitudes are the most reliable for extended photometry. As we show in Figure \ref{comp_mags}, the difference between the $r$-band ``modelMag'' and ``auto'' magnitudes is generally not significant, and there is no global systematic offsets between the magnitudes of extended Type 1 AGN and point sources. However, we caution that individual sources could have discrepant  ``modelMag'' magnitudes, and the effective $R-W1$ colors calculated here may not always reflect the true $R-W1$ colors for Type 1 AGN with extended photometry.

We convert $r$ from SDSS filters in the AB system to the Bessell $R$ bandpass \citep{bessell} in the Vega system to be consistent with many previous red quasar studies \citep[e.g.,][]{brusa1,brusa15,georgakakis,banerji1}, following the formulae provided in \citet{blanton}:
\begin{equation}
R_{\rm AB} = r - 0.0576 - 0.3718((r - i) - 0.2589)
\end{equation}
\begin{equation}
R_{\rm Vega} = R_{\rm AB} -0.21,
\end{equation}
where the latter equation converts Bessell $R$ from the AB to the Vega magnitude system. For $W1$, we use the instrumental profile-fit photometry magnitude, ``w1mpro'', in the {\it WISE} All-Sky Source catalog.

\subsection{Spectral Evolution with Redshift}\label{sedz}
\begin{figure}
{\includegraphics[scale=0.4,angle=90]{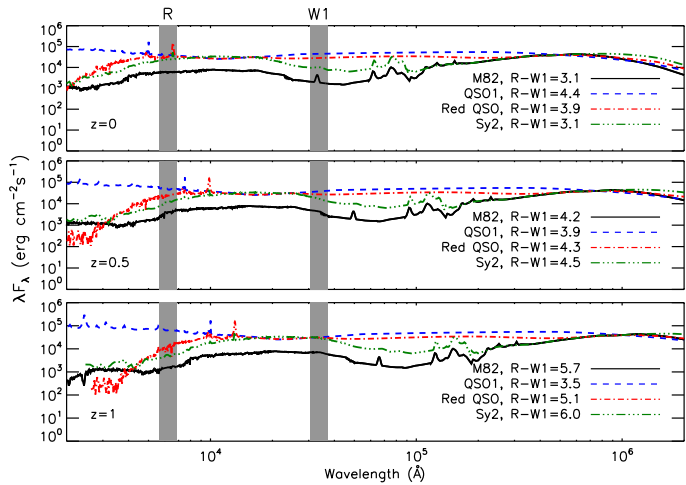}}
\caption[]{\label{seds1} SWIRE template spectra of a Type 1 quasar (blue dashed), reddened broad-lined quasar (red dot-dashed), Seyfert 2 galaxy (moderate luminosity narrow-line AGN; green dot-dot-dashed) and the M82 starburst galaxy (black solid) for $z$=0, 0.5, 1 \citep{polletta}. The grey-shaded regions mark the $R$ and $W1$ passbands. Calculated R-W1 colors for each object at the various redshifts are noted in the legend. Red $R-W1$ ($>$4) colors is  an excellent discriminator between obscured and unobscured accretion beyond the local Universe, especially at $z>1$.}
\end{figure}

To illustrate how the $R$ and $W1$ bands overlap spectral regions that can provide clues as to the type of object being observed, we use the SWIRE template library of \citet{polletta} to plot the spectra for a Type 1 quasar, a reddened quasar \citep[based on the FIRST red quasar J013435.7-093102,][]{gregg}, and a Seyfert 2 galaxy (Sy2, moderate luminosity obscured AGN) at different redshifts, with the spectrum of starburst galaxy M82 shown for reference, in Figure \ref{seds1}; the spectra are normalized at 50 $\mu$m to ease comparison. Here, the grey bands indicate the wavelength coverage of the $R$ and $W1$ bands, respectively. We also note the calculated $R-W1$ values for each spectrum in the legend. As redshift increases, the $R$- band flux becomes increasingly attenuated for the red quasar and Sy2 spectra while the $W1$-band flux remains relatively steady, reflected in an enhancement in the $R-W1$ color.

\begin{figure}
{\includegraphics[scale=0.4,angle=90]{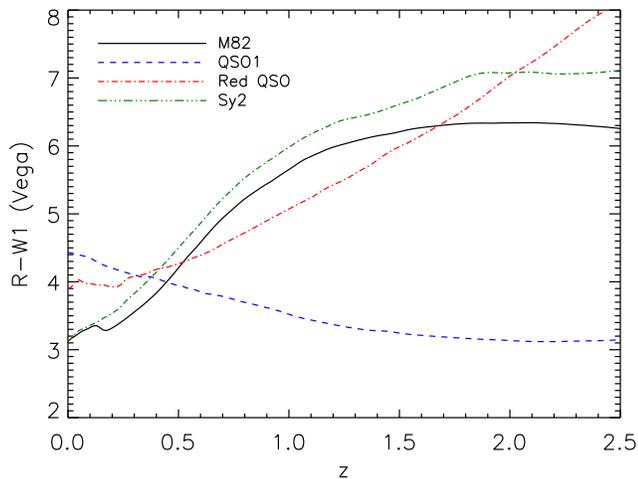}}
\caption[]{\label{r-w1_v_z} Calculated $R-W1$ values from the \citet{polletta} templates as a function of redshift for various classes of objects. At $z>0.5$ the divergence between unobscured quasars and obscured AGN (both reddened quasars and Seyfert 2 galaxies) is appreciable, though starburst galaxies also have red colors at these distances. X-ray emission can then differentiate between obscured accretion and non-active galaxies.}
\end{figure}

We demonstrate this trend of optical reddening with distance further in Figure \ref{r-w1_v_z} where we plot the fiducial $R-W1$ values, calculated from the \citet{polletta} templates, as a function of redshift. While $R-W1$ increases with redshift for the reddened quasar and Sy2 galaxy templates ($R-W1>4-8$), the Type 1 quasar remains approximately constant and is predominantly bluer ($R-W1 \sim 3-3.9$). The exception of redder $R-W1$ colors being associated with obscured accretion is at $z \sim0$, where both reddened quasars and Sy2s have bluer $R-W1$ colors ($R-W1<4$). 

\begin{figure}
{\includegraphics[scale=0.4,angle=90]{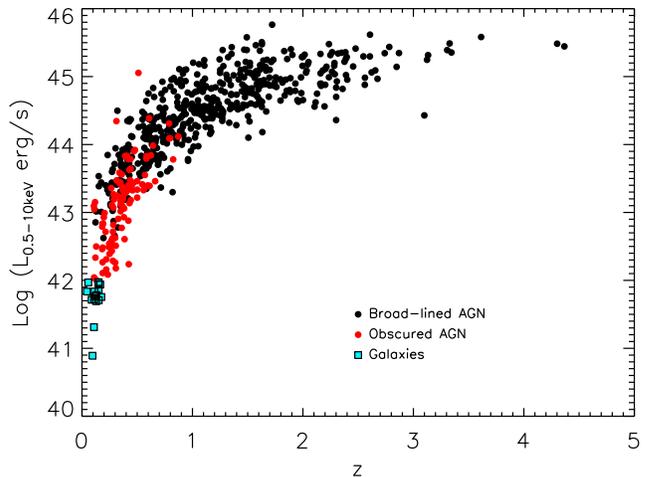}}
\caption[]{\label{z_lum} Observed-frame X-ray luminosity (0.5-10 keV) as a function of redshift for the extragalactic sources in the $R-W1$ sample, with different classes of object marked as in the legend. Here, ``obscured AGN'' refer to systems optically classified as galaxies yet have X-ray luminosities exceeding 10$^{42}$ erg s$^{-1}$ while ``galaxies'' are sources with X-ray luminosites below this threshold. Most (68\%) of the highest luminosity AGN ($>10^{44}$ erg s$^{-1}$) are at $z>1$, where $R-W1$ is a robust discriminator between the obscured and unobscured AGN populations. }
\end{figure}

As Figure \ref{z_lum} illustrates, the higher luminosity AGN in our sample ($>10^{44}$ erg s$^{-1}$) are predominantly at $z>1$ (due to the large-area and shallow flux limit of Stripe 82X), such that $R-W1$ is an excellent discriminator between the obscured and unobscured quasar population. We note however, that non-active galaxies also become redder as redshift increases, meaning that sources selected based on just $R-W1$ colors from a parent optical or infrared sample will include a non-negligible fraction of non-active galaxies. X-ray emission then becomes an essential clue for separating AGN from galaxies hosting dormant black holes.

Our focus is then on how the combination of X-ray, optical, and infrared data can be combined to open a new window into obscured black hole growth by examining the $R-W1$ colors of X-ray selected sources to identify potentially obscured AGN. In addition to current wide-area X-ray surveys like Stripe 82X, {\it XMM}-XXL,  and the {\it XMM}-Serendipitous Survey, {\it eROSITA} will launch in 2017 \citep{erosita}, surveying the full-sky from 0.3-10 keV with an angular resolution of $<15^{\prime\prime}$ on-axis, and will detect millions of AGN candidates \citep{merloni}. Though the results here may not be applicable to optically-selected or infrared-selected samples, or objects found in very deep X-ray surveys, and the Stripe 82X sample is not complete (i.e., only $\sim$59\% of the X-ray objects with SDSS and {\it WISE} counterparts within the magnitude limits discussed in Section \ref{samp} are currently identified via spectroscopic redshifts), the trends we report below are likely to represent the AGN demographics of the sources initially discovered in current and future shallow X-ray surveys, where $R-W1$ can easily be calculated over most of the sky thanks to existing facilities like {\it WISE} and PanSTARRS.

\section{Results}

\subsection{$R-W1$ Sample} 
\begin{figure}
{\includegraphics[scale=0.4,angle=90]{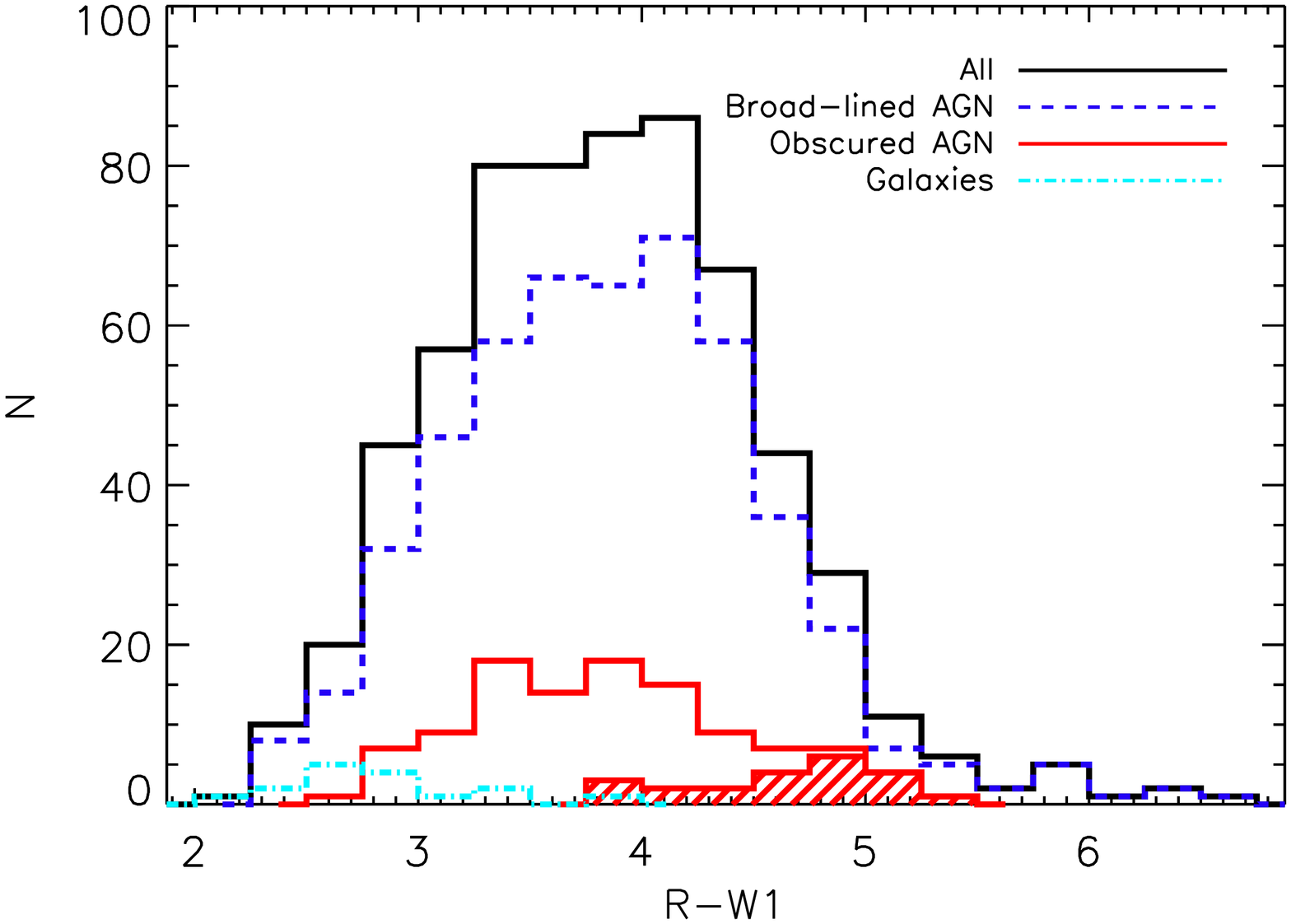}}
{\includegraphics[scale=0.4,angle=90]{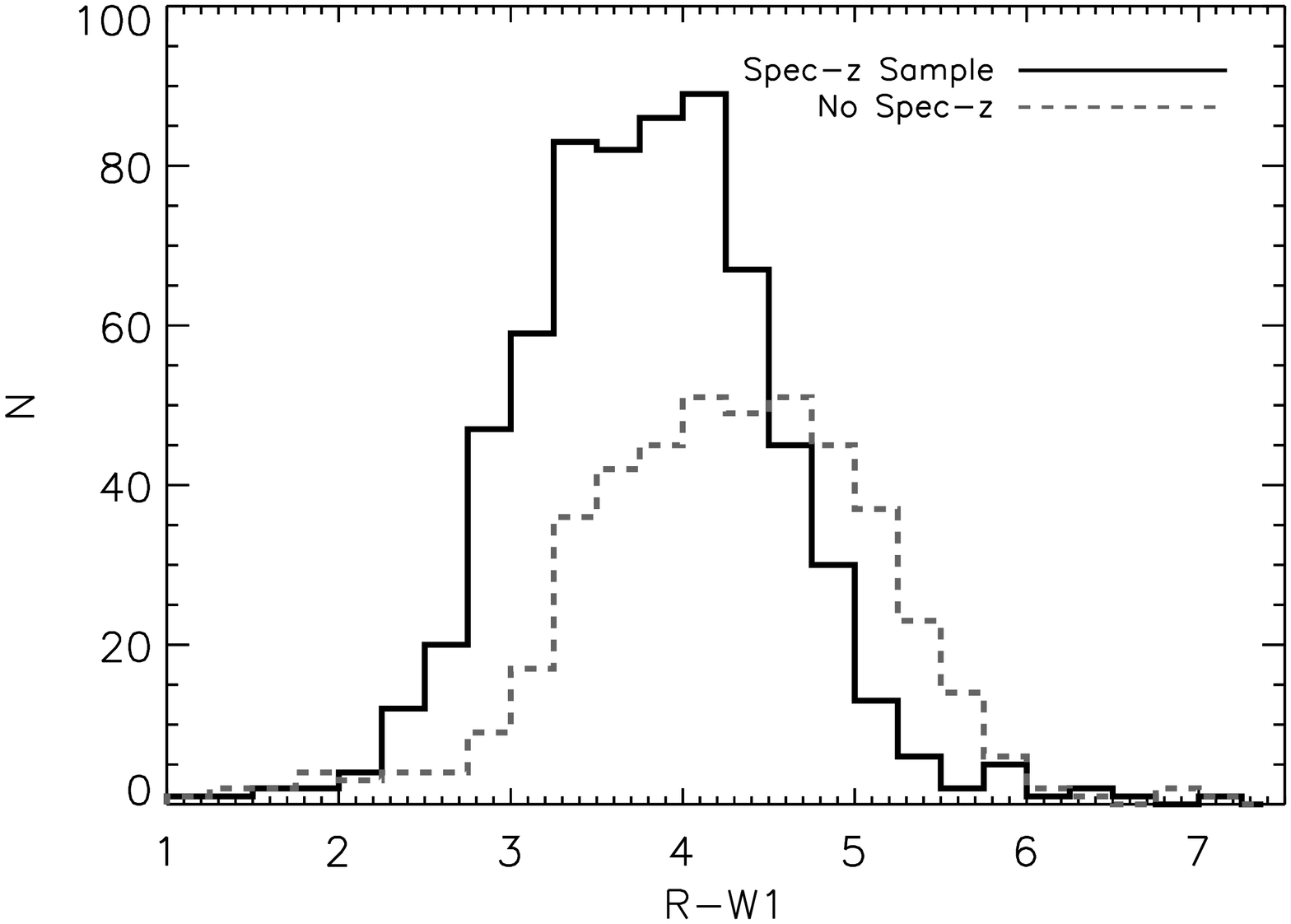}}
\caption[]{\label{s82x_samp} {\it Top}: $R-W1$ distribution for extragalactic objects in the Stripe 82X sample. The red shaded histogram indicates the obscured AGN at $z>0.5$, where BPT diagnostics using H$\alpha$ and [NII] are untenable in observed frame optical spectra. While most of the sample is broad-lined AGN, they have a wide range of $R-W1$ colors. {\it Bottom}: Comparison of the  $R-W1$ distribution for the spectroscopic Stripe 82X  sample (black line) and the X-ray sources lacking redshifts (dashed grey line). The unidentified X-ray sources have systematically redder $R-W1$ colors, indicating that there are obscured AGN still to be identified.}
\end{figure}

In Figure \ref{s82x_samp} (top), we plot the $R-W1$ distribution of the 631 extragalactic Stripe 82X sources in our sample, highlighting broad-lined AGN; obscured AGN, which are categorized as ``galaxies'' based on their optical spectra but have X-ray luminosities consistent with accretion onto a supermassive black hole; and galaxies, which have both optical signatures and X-ray luminosities consistent with non-active galaxies. We also highlight in figure \ref{s82x_samp} (top) the subset of obscured AGN at $z>0.5$, which is the redshift range where BPT diagnostics using H$\alpha$ and [NII] become unfeasible for observed-frame optical spectra (see Section \ref{opt_gal} for more discussion); most of these obscured AGN have red $R-W1$ ($>$4) colors. We note that while a majority of this Stripe 82X sample consists of broad-lined AGN, they have a range of $R-W1$ colors, rather than being predominantly blue, indicating the presence of a significant fraction of reddened quasars in our sample.

In Figure \ref{s82x_samp}, we compare the $R-W1$ distribution of the Stripe 82X sources lacking spectroscopic redshifts, but meeting the SDSS magnitude and $W1$ SNR and positional cuts detailed above, with the spectroscopic {\it R-W1} sample considered in this work, representing 59\% of the X-ray, SDSS, and {\it WISE} sources have spectroscopic redshifts. The X-ray sources without spectra are systematically redder than the objects identified thus far, implying that there are obscured supermassive black holes at $z>0.5$ in Stripe 82X yet to be confirmed.

\subsection{\label{stellar_locus} Differentiating Galactic from Extragalactic Objects}
\begin{figure}
{\includegraphics[scale=0.4,angle=90]{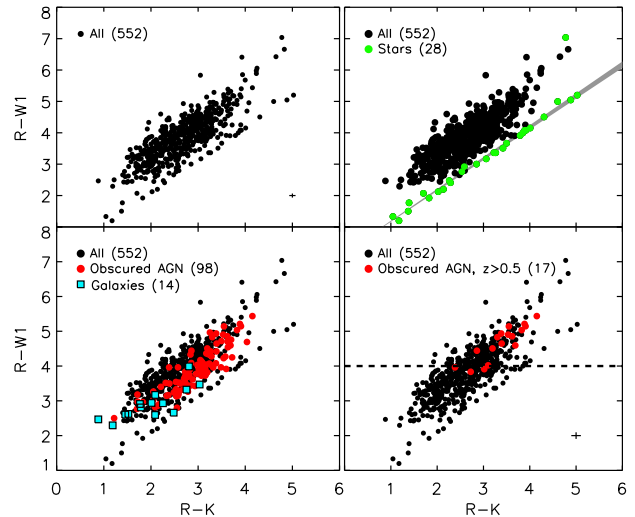}}
\caption[]{\label{r-w1_v_r-k_a} Stripe 82 X-ray sources with spectroscopic redshifts and detections in the UKIDSS $K$ and {\it WISE} $W1$ bands (black circles), with different classes highlighted by the colors as indicated in the legend: the $R-W1$ (Vega) diagnostic is well correlated with the established $R-K$ (Vega) color. Stars follow a well-defined path in $R-K$ vs. $R-W1$ space ({\it top right}), with the best-fit relation, including the error on the slope, overplotted. Normal galaxies ($L_{\rm x} < 10^{42}$ erg s$^{-1}$; no broad lines in optical spectrum) tend to have bluer colors while obscured AGN ($L_{\rm x} > 10^{42}$ erg s$^{-1}$; no broad lines in optical spectrum) have a range of colors ({\it bottom left}). Optically obscured AGN at $z>0.5$ (the redshift at which BPT diagnostics using H$\alpha$ in observed-frame optical spectra are not available) have red colors ({\it bottom right}); here, the dashed line indicates our proposed $R-W1>4$ color cut to separate optically elusive AGN and XBONGs from non-active galaxies in X-ray samples; reddened broad-lined AGN and some stars also lie above this line. The median error bars of the $R-W1$ and $R-K$ colors for the full sample and the obscured AGN population at $z>0.5$ are shown in the top left and bottom right panels, respectively; the error bars for the other sub-populations are smaller than the symbol size and are thus not plotted.}
\end{figure}

\begin{figure}
{\includegraphics[scale=0.4,angle=90]{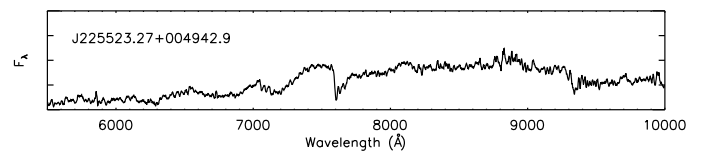}}
\caption[]{\label{mdwarf} Palomar DoubleSpec spectrum of the M-dwarf star which is the outlier from the stellar locus plotted in \ref{r-w1_v_r-k_a}; the name in the legend is based on the X-ray coordinates \citep[see][for the X-ray catalog]{me2}. The spectrum has not been flux calibrated nor corrected for telluric absorption.}
\end{figure}

Figure \ref{r-w1_v_r-k_a} shows that $R-W1$ is well correlated with $R-K$ for the subset of our sample that have $W1$ and $K$ detections. Stars generally follow a well-defined track in $R-K$ vs. $R-W1$ color space (top right panel of Figure \ref{r-w1_v_r-k_a}). There is one obvious stellar outlier at $R-W1>6$, which is an M-dwarf detected by our Palomar DoubleSpec follow-up observing campaign (see Figure \ref{mdwarf}); however, we note that other M-dwarfs follow the stellar sequence. We fit the relation between $R-W1$ and $R-K$ for the stars, omitting the one extreme outlier, finding
\begin{equation}
R-W1 = 0.998 (\pm0.02) \times (R-K) + 0.18, 
\end{equation}
with a correlation coefficient of 0.997. This relationship, overplotted in Figure \ref{r-w1_v_r-k_a}, can be used to distinguish between Galactic and extragalactic X-ray sources in the absence of spectroscopic data or when significant {\it WISE} $W3$ detections are lacking, causing identifications based on the {\it WISE} color diagnostic plot \citep[i.e., $W1-W2$ versus $W2-W3$;][]{wise} to be unfeasible. Additionally, this trend provides a more efficient method for targeting specific classes of objects in follow-up campaigns over using a simple $R-K$ or $R-W1$ color cut, should $K$-band coverage be available.

%\FloatBarrier
\subsection{\label{opt_gal} Optical Galaxies}
In our sample, objects spectroscopically classified as galaxies are sources that lack broad emission lines. These objects can be obscured AGN (either narrow-lined sources or objects where host galaxy dilution obliterates weak signatures of supermassive black hole accretion), blazars which have featureless optical spectra, or galaxies hosting dormant black holes. Since X-rays pierce through the dust that attenuates optical emission and do not suffer dilution effects, such emission cleanly separates active from inactive galaxies when the X-ray luminosity exceeds 10$^{42}$ erg s$^{-1}$ \citep[e.g.,][]{brandt,brandt2015}. In the bottom left-hand panel of Figure \ref{r-w1_v_r-k_a}, we use the X-ray emission to split the optical galaxy sample into AGN ($L_{\rm x} > 10^{42}$ erg s$^{-1}$) and normal galaxies ($L_{\rm x} < 10^{42}$ erg s$^{-1}$);\footnote{Here, L$_{\rm x}$ refers to the full-band luminosity if the full-band X-ray flux is significant. Otherwise, we base the AGN cut on the X-ray luminosity exceeding 10$^{42}$ erg s$^{-1}$ in either the soft- or hard-band.} we caution however that the ``normal galaxies'' in this subsample can include heavily obscured AGN, where the observed X-ray emission is severely extincted by Compton-thick levels of obscuration, or low-luminosity AGN. 

While non-broad-lined AGN have a range of colors, normal galaxies tend to have bluer colors ($R-W1 < 4$).  This latter effect is mostly induced by the flux-limited nature of our sample: due to the relatively short exposure times for the majority of the X-ray observations, these galaxies are local, spanning a redshift range of $0.04 < z < 0.51$ where the observed spectrum is bluer than it is at higher redshifts \citep[Figures \ref{seds1}-\ref{r-w1_v_z}; see also][]{brusa2010}. Beyond $z>0.5$, most of the AGN optically classified as galaxies have red colors (Figure \ref{r-w1_v_r-k_a}, bottom right), as expected based on the spectral templates.

\begin{figure}
\begin{center}
\includegraphics[scale=0.5]{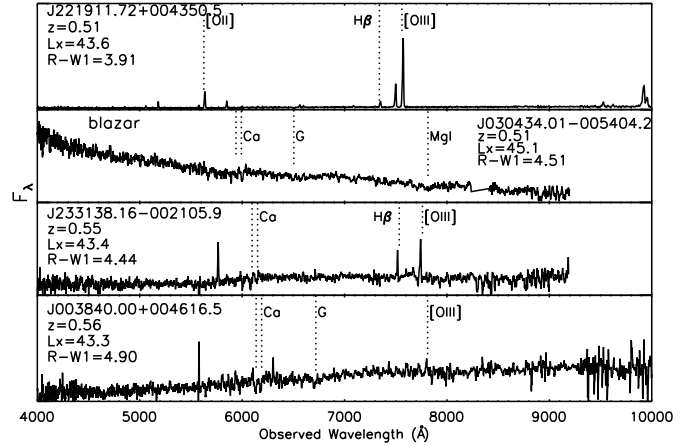}
\caption{\label{nl_agn1}SDSS spectra of optically classified ``galaxies'' that have X-ray luminosities consistent with AGN. The marked transitions indicate how the redshift was determined. The names indicated in the legend are based on the X-ray coordinates. $L_{\rm x}$ is measured in the observed 0.5-10 keV band (except for the top source, where the full-band flux was not significant and the reported luminosity is in the 2-10 keV band) and is reported in log space. The object in the first panel can be classified as an AGN on the strength of the [OIII] 5007\AA\  line while the source in the second panel is a blazar.}
\end{center}
\end{figure}

\begin{figure}
\begin{center}
\includegraphics[scale=0.5]{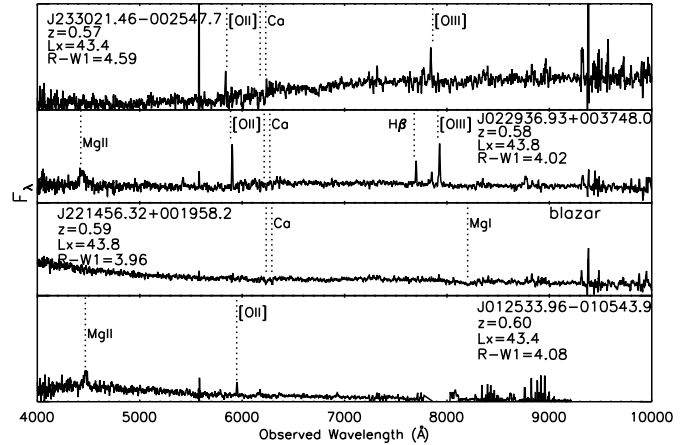}
\caption{\label{nl_agn2}SDSS spectra of optically classified ``galaxies'' that have X-ray luminosities consistent with AGN ({\it continued from Figure \ref{nl_agn1}}). While the objects in the second and fourth panels have a broader MgII emission line compared with the other species, SDSS classified  these sources as ``galaxies.''  The object in the third panel is a blazar.}
\end{center}
\end{figure}

\begin{figure}
\begin{center}
\includegraphics[scale=0.5]{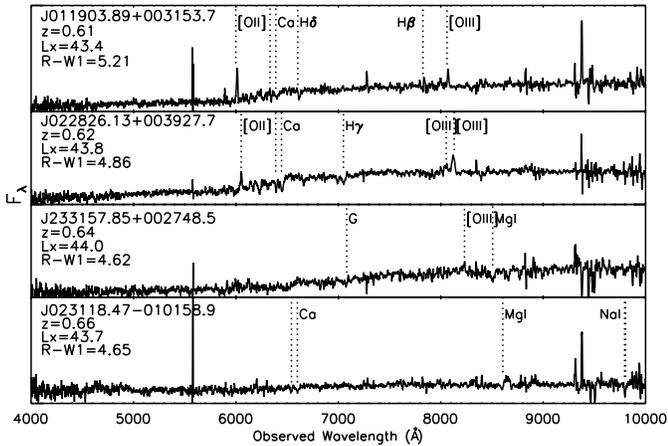}
\caption{\label{nl_agn3}SDSS spectra of optically classified ``galaxies'' that have X-ray luminosities consistent with AGN ({\it continued from Figure \ref{nl_agn1}}).}
\end{center}
\end{figure}

\begin{figure}
\begin{center}
\includegraphics[scale=0.5]{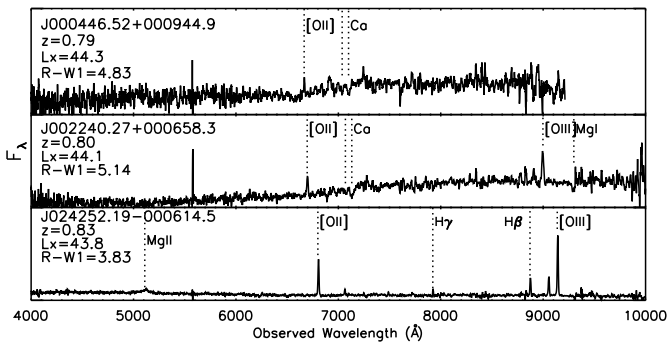}
\caption{\label{nl_agn4}SDSS spectra of optically classified ``galaxies'' that have X-ray luminosities consistent with AGN ({\it continued from Figure \ref{nl_agn1}}).  While the object in the bottom panel has a broader MgII emission line compared with the other species, SDSS classified this source as a ``galaxy.'' }
\end{center}
\end{figure}

\begin{figure}
\begin{center}
\includegraphics[scale=0.5]{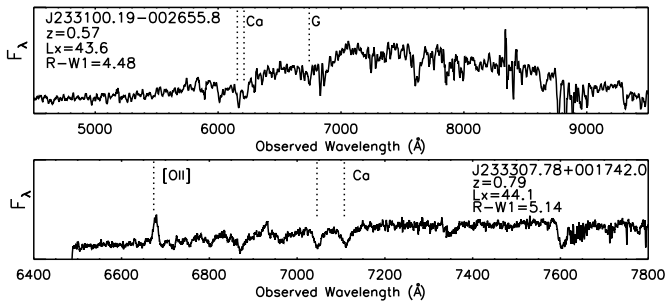}
\caption{\label{nl_agn5} ({\it Top}) WIYN HYDRA spectrum from our follow-up campaign (not flux calibrated nor telluric-corrected) and ({\it bottom}) DEEP 2 spectrum of optically classified ``galaxies'' that have X-ray luminosities consistent with AGN ({\it continued from FIgure \ref{nl_agn1}}).}
\end{center}
\end{figure}

In Figures \ref{nl_agn1} - \ref{nl_agn5}, we plot the optical spectra of most of these $z>0.5$ obscured AGN,\footnote{Two objects were identified via catalogs released from the PRIMUS survey \citep{coil}, whose spectra are not public.} including two AGN in the $R-W1$ sample that are not part of the UKIDSS subsample, since this is a redshift where $R-W1$ is a good indicator of obscured accretion and BPT diagnostics using observed-frame H$\alpha$ and [NII] are unfeasible. Two of the objects are blazars, based on both their largely featureless optical spectra (J030434.01-005404.2 and J221456.32+001958.2) and detections in the FIRST radio survey \citep{first,me2}; the source in the top panel of Figure \ref{nl_agn1} (J221911.72+004350.5) can be classified as an AGN based on the strength of the [OIII] 5007 \AA\  line; and three have broader MgII emission lines than the other species which were not well-fit by the SDSS pipeline and thus not categorized as AGN (J022936.93+003748.0, J024252.19-000614.5, and J012533.96-010543.9). 

The remaining 11 sources, all of which have $R-W1 > 4$ colors, can be categorized as ``optically elusive AGN'', which have emission lines consistent with star-forming galaxies or LINERs \citep{maiolino,caccianiga,smith}, and X-ray Bright Optically Normal Galaxies \citep[XBONGs;][]{fiore2000,cocchia}, where the optical spectrum shows absorption features consistent with a passive galaxy. The optical normalcy of elusive AGN and XBONGs may be caused by dilution from host galaxy starlight \citep[e.g.,][]{moran,caccianiga,georgantopoulos,civano2007}, attenuation by dust \citep[e.g.,][]{comastri2002,smith,rigby}, or perhaps radiatively inefficient accretion flows \citep{yuan} or more speculatively, the ``switching-on'' of the AGN, where the narrow-line region has yet to be photoionized \citep{schawinski}. 

We propose that a color cut of $R-W1>4$ in X-ray samples from shallow surveys efficiently separates optically elusive AGN and XBONGs from non-active galaxies at $z>0.5$; in deeper X-ray surveys, like the {\it Chandra} Deep Field South \citep{giacconi,xue}, a higher fraction of non-active galaxies are detected at $z>0.5$ \cite[e.g.,][]{lehmer} which would also have $R-W1>4$ colors (see Figure \ref{seds1}-\ref{r-w1_v_z}). We note that this cut on the Stripe 82X data does not select optically elusive AGN/XBONGs exclusively, but also includes reddened broad-lined quasars and a handful of X-ray emitting stars. Using just $R-W1$, the percentage of stellar contaminants to all active galaxies in this color-cut is $\sim$3\%; when $R-W1>4$ is used in conjunction with $R-K$ and the relationship introduced in Section \ref{stellar_locus}, this stellar contamination drops to under 0.9\%. None of the non-active galaxies have $R-W1$ colors exceeding 4.

%\FloatBarrier
\subsection{Trends with Observed X-ray Luminosity}
\begin{figure}[ht]
{\includegraphics[scale=0.4,angle=90]{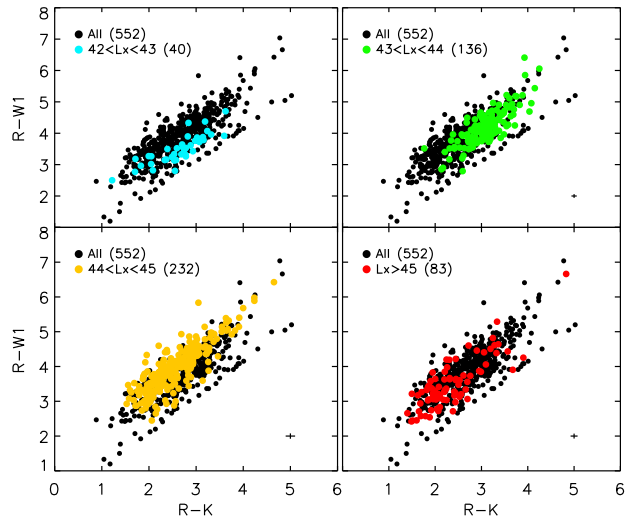}}
\caption[]{\label{r-w1_v_r-k_b} Same as Figure \ref{r-w1_v_r-k_a}, but with highlighted sources indicating various observed X-ray luminosity bins in the full-band (0.5-10 keV). Moderate luminosity AGN, $10^{43}$ erg s$^{-1} <$ $L_{\rm 0.5-10keV} < 10^{44}$ erg s$^{-1}$ ({\it top right}), tend to have redder colors, while the lowest luminosity ($10^{42}$ erg s$^{-1} <$ $L_{\rm 0.5-10keV} <10^{43}$ erg s$^{-1}$, {\it top left}) and highest luminosity AGN ($L_{\rm 0.5-10keV} >10^{45}$ erg s$^{-1}$, {\it bottom right}) have bluer colors. This is further illustrated in Figure \ref{r-w1_hist}. The median error bars for the separate luminosity bins are plotted in each panel with the exception of the lowest luminosity AGN bin since the error bars are smaller than the symbol size.}
\end{figure}

\begin{figure}[ht]
{\includegraphics[scale=0.4,angle=90]{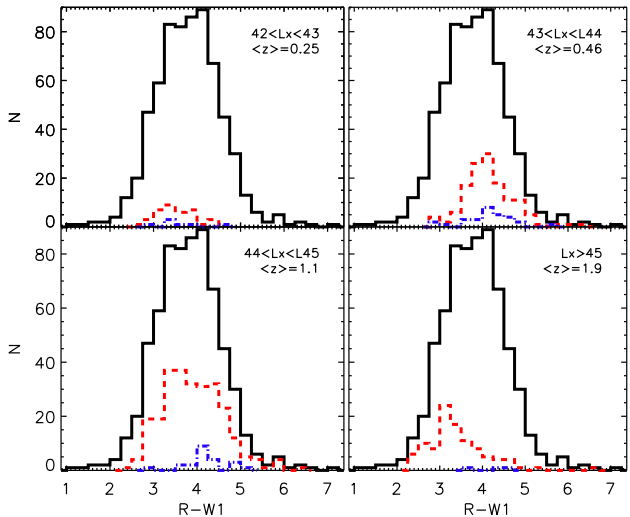}}
\caption[]{\label{r-w1_hist} Distribution of $R-W1$ for the Stripe82X sample (black solid line), with each panel focusing on different observed full-band X-ray luminosity ranges; the mean redshift for each sub-sample is noted in the legend. The red dashed line denotes objects with spectroscopic redshifts from pre-existing optical surveys and the blue dot-dashed line mark the sources identified by our dedicated follow-up campaigns. A trend in observed X-ray luminosity is apparent, where the lowest luminosities have bluer colors ({\it top left}), moderate luminosity AGN have redder colors ({\it top right}), moderately high luminosity AGN have a range of colors ({\it bottom left}), and high luminosity AGN are predominantly blue ({\it bottom right}); however these luminosity bins are incomplete with respect to the parent Stripe 82X sample from which they are culled. The moderately high luminosity AGN identified via our spectroscopic follow-up tend to have redder colors, while the newly identified AGN in other luminosity ranges more closely follow the color distribution of the AGN identified by optical surveys.}
\end{figure}

\begin{figure}[ht]
{\includegraphics[scale=0.4,angle=90]{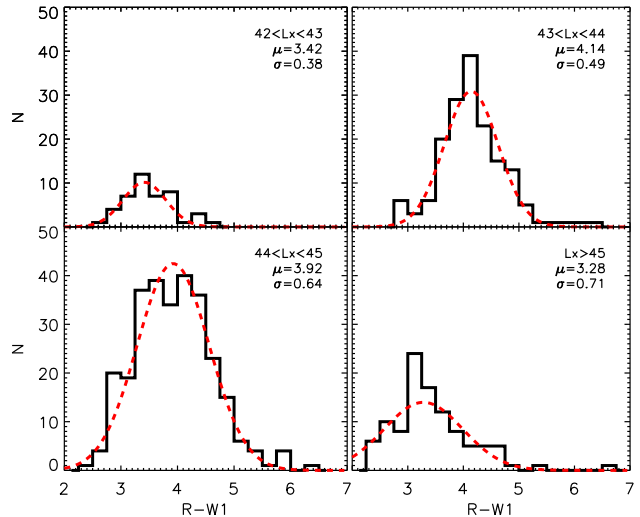}}
\caption[]{\label{r-w1_hist_fit} Distribution of $R-W1$ for specific luminosity ranges, where the distribution is fitted by a Gaussian, with fit parameters noted in the legend. The various luminosity bins have different mean $R-W1$ values, though the width of the distributions tend to be rather wide. }
\end{figure}

Here we explore $R-W1$ and $R-K$ colors of the Stripe 82X AGN for different observed full-band X-ray luminosity bins, to test whether there is an association between optical reddening and observed X-ray emission, though we caution that these sub-samples are incomplete. As seen in Figures \ref{r-w1_v_r-k_b}-\ref{r-w1_hist_fit}, AGN with different observed X-ray luminosities do occupy various regions of optical-infrared parameter space. The lowest luminosity AGN ($10^{42}$ erg s$^{-1} <$ $L_{\rm 0.5-10keV} <10^{43}$ erg s$^{-1}$) have bluer colors, as expected for Seyfert 2 galaxies at $z\sim0$ (Figure \ref{seds1}). The color trend shifts noticeably redward at moderate X-ray luminosities ($10^{43}$ erg s$^{-1} <$ $L_{\rm 0.5-10keV} < 10^{44}$ erg s$^{-1}$), where the redshifts of this population extends beyond the local Universe: red $R-W1$ and $R-K$ colors here and at higher luminosities indicate obscured accretion. At higher X-ray luminosities ($10^{44}$ erg s$^{-1} <$ $L_{\rm 0.5-10keV} < 10^{45}$ erg s$^{-1}$), the distribution spans a full range of colors. The highest luminosity AGN ($L_{\rm 0.5-10keV} >10^{45}$ erg s$^{-1}$) are instead predominantly blue, but due to the limited spectroscopic completeness of the sample, it is unclear whether this trend is a selection effect or a physical one (see Section 5). 

In Figure \ref{r-w1_hist}, we differentiate between sources that have spectroscopic redshifts from optical surveys (red dashed line) and those with redshifts from our follow-up campaign (blue dot-dashed line). We caution that only $\sim$12\% of our sample have these latter redshifts and very few of these fall in the lowest and highest luminosity bins ($10^{42}$ erg s$^{-1} <$ $L_{\rm 0.5-10keV} <10^{43}$ erg s$^{-1}$ and $L_{\rm 0.5-10keV} >10^{45}$ erg s$^{-1}$, respectively). The AGN we confirmed from our follow-up work tend to have redder $R-W1$ colors at luminosities $10^{44}$ erg s$^{-1} <$ $L_{\rm 0.5-10keV} < 10^{45}$ erg s$^{-1}$, while AGN with existing redshifts at the same luminosities have a broader range of colors. Hence our spectroscopic campaign is recovering obscured quasars not identified by optical surveys since X-ray selection is sensitive to objects made fainter by dust that are thus missed by optical selection.

We quantify the luminosity trends in Figure \ref{r-w1_hist_fit} by fitting a Gaussian to each luminosity bin to determine the mean and spread of the $R-W1$ distributions. We find that the mean $R-W1$ color varies among the luminosity bins, with values of 3.42 ($\sigma=0.38$), 4.14 ($\sigma=0.49$), 3.92 ($\sigma=0.64$), and 3.28 ($\sigma=0.71$) from the lowest to highest luminosities. However, the width of the distributions are quite wide due to the large range of colors sampled, suggesting that the differences in average $R-W1$ color are not statistically significant. Additionally, when dividing the $10^{43}$ erg s$^{-1} <$ $L_{\rm 0.5-10keV} < 10^{44}$ erg s$^{-1}$ and $10^{44}$ erg s$^{-1} <$ $L_{\rm 0.5-10keV} < 10^{45}$ erg s$^{-1}$ sub-populations into finer 0.5 dex luminosity bins, the width of the $R-W1$ distribution does not change, suggesting that the scatter is not induced by sampling a relatively larger population.

%\FloatBarrier

\section{\label{disc} Discussion}
Taken at face-value, our results that moderate luminosity AGN have red colors while the highest luminosity AGN have bluer colors indicate that the former are obscured while the latter are unobscured. Similar results have been found in past X-ray surveys that used gas column density ($N_{\rm H}$) as an indicator of obscuration \citep{ueda,treister,hasinger,merloni2014,ueda2014,lanzuisi,buchner}. 

However, the bluer colors of the most luminous AGN seems to be at odds with deeper, smaller-area X-ray surveys. \citet{donley} and \citet{eckart} found a trend between X-ray luminosity and IRAC colors in the $\sim$2 deg$^2$ COSMOS and Serendipitous Extragalactic X-ray Source Identification surveys, respectively, with redder mid-infrared emission associated with more luminous AGN. Similarly, using infrared and X-ray data from the 0.3 deg$^2$ Extended {\it Chandra} Deep Field South survey \citep{lehmer,virani}, \citet{cardamone} demonstrated that X-ray luminosity trends with redder infrared emission below 8$\mu$m. However, since both these surveys cover a much smaller area than Stripe 82X, only a handful of the most X-ray luminous AGN ($>10^{45}$ erg s$^{-1}$) are found, while 109 are used in this analysis. Thus, we are able to extend the lever arm of this comparison to higher luminosities than formerly explored, albeit with the caveat that this association only holds for the X-ray sources we have thus far identified with spectroscopic redshifts. Also, the optical-infrared colors we use here sample different portions of the SED than the infrared colors used in the past studies, and we work with observed, non-absorption corrected luminosities rather than rest-frame, absorption corrected luminosities, as the previous studies used, since these corrections are uncertain.

There may be a population of $L_{\rm 0.5-10keV} > 10^{45}$ erg s$^{-1}$ AGN from the parent Stripe 82 X-ray source list that are not yet identified since they lack spectroscopic redshifts, or indeed, even optical counterparts. For instance, the {\it XMM}-COSMOS survey finds that 20\% of the $L_{\rm X} > 10^{45}$ erg s$^{-1}$ quasar population is obscured \citep{brusa2010, merloni2014}: though fewer of the rare high-luminosity quasars are found in this 2.2 deg$^2$ survey compared with wider-area surveys like Stripe 82X, the X-ray identification rate via spectroscopic and photometric redshifts is near 100\%, allowing elusive AGN to be readily identified. In addition to our optical campaign to increase the spectroscopic completeness of Stripe 82X, and the calculation of photometric redshifts via SED fitting, we have a dedicated follow-up near-infrared program to target obscured AGN candidates specifically, some of which lack optical counterparts. Preliminary results from these observing campaigns have revealed several $L_{\rm 0.5-10keV} \sim 10^{44} -  10^{45}$ erg s$^{-1}$ quasars that are heavily reddened (LaMassa et al. {\it in prep.}), but future observations are necessary to test whether this population is significant and surpasses that of the unobscured AGN we have identified thus far.

\subsection{$W1-W2$-selected AGN}
$R-W1$ provides an indication of how much the optical emission in AGN is attenuated, causing the spectrum to be reddened compared with the blue continuum observed in unobscured sources. How does this compare with AGN selected on the basis of their mid-infrared colors only? \citet{stern12} introduced a $W1-W2>0.8$ color cut to select AGN for $W2$ magnitudes brighter than 15, calibrated on {\it Spitzer}-selected AGN \citep{stern2005} in the COSMOS field. \citet{assef} pushed the $W1-W2$ color cut down to fainter magnitudes in the Bo\"otes field, which has deeper {\it WISE} coverage than COSMOS, where they found that $W1-W2>0.662 \times$exp$(0.232\times(W2-13.97)^2)$ for $W2<17.11$ has a 90\% reliability in selecting AGN they identified via SED fitting. Here, we define $W1-W2$ selected AGN as sources with $W1-W2>0.8$ for $W2<15$ and $W1-W2>0.662 \times$exp$(0.232\times(W2-13.97)^2)$ for $15<W2<17.11$; we restrict our Stripe 82X spectroscopic sample to the sources that have $W2$ SNRs $\geq$5 for this analysis, resulting in 621 objects.

\begin{figure}[ht]
{\includegraphics[scale=0.4,angle=90]{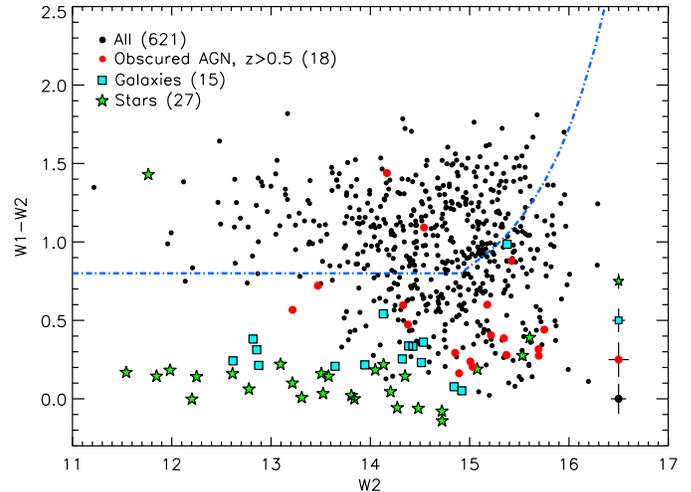}}
\caption[]{\label{w1w2_class} $W1-W2$ versus $W2$ for sources in the Stripe 82X sample with $W2$ SNR $>$5, where objects above the blue dot-dashed line are AGN as defined by the \citet{stern12} and \citet{assef} classifications. Optically elusive AGN and XBONGs at $z>0.5$, which are mostly identified with a $R-W1>4$ color cut, are largely not selected as AGN by the $W1-W2$ metric. The black data points represent Type 1 AGN at all redshifts and obscured AGN at $z<0.5$. Median error bars for each population are shown on the right side of the panel.}
\end{figure}

In Figure \ref{w1w2_class}, we plot $W1-W2$ versus $W2$ and highlight where the stars, galaxies ($L_{\rm 0.5-10keV}< 10^{42}$ erg s$^{-1}$), and optically elusive AGN and XBONGs at $z>0.5$ ($L_{\rm 0.5-10keV}> 10^{42}$ erg s$^{-1}$) live in this parameter space. Most of the obscured AGN, which are selected with a $R-W1>4$ color-cut, do not populate the $W1-W2$ AGN locus (above the blue dot-dashed line). A similar conclusion was reached by \citet{barmby} and \citet{brusa2010} who reported that a large fraction of obscured AGN at $z>1$ lie outside the \citet{stern2005} IRAC AGN color-track, while the \citet{lacy2004} IRAC color-selections were more likely to recover obscured AGN, and by Menzel et al. ({\it submitted}) who found that obscured, host-galaxy dominated, and optically elusive AGN in XXL-N do not meet the $W1-W2$ color cut. Additionally, though stars follow a well-defined track in the $R-K$ versus $R-W1$ parameter space, no clear differentiation between Galactic and extragalactic objects based on $W1-W2$ color is apparent. While stars generally have $W1-W2$ colors around zero, so too do some galaxies and obscured AGN.

\begin{figure}[ht]
{\includegraphics[scale=0.4,angle=90]{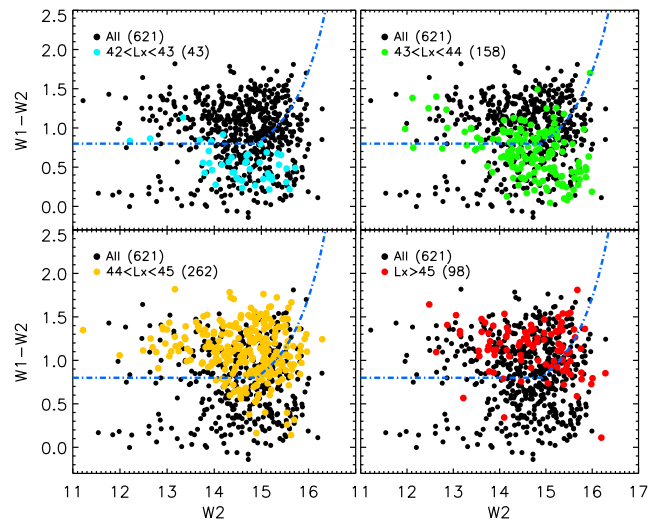}}
\caption[]{\label{w1w2_lum} $W2$ versus $W1-W2$ for the Stripe 82X sample with the different X-ray luminosity populations highlighted. Mid-infrared AGN selection works best at identifying luminous AGN, with over $\sim$66\% of the $L_{\rm 0.5-10keV}>10^{44}$ erg s$^{-1}$ sources lying within the $W1-W2$ AGN locus.}
\end{figure}

Identifying AGN based on infrared-only colors has the highest reliability at higher flux limits \citep[e.g.,][]{barmby,donley2008,cardamone,eckart,donley,mendez}. As we show in Figure \ref{w1w2_lum}, the efficacy of $W1-W2$ in selecting X-ray identified AGN increases with luminosity. While 14\% of the lowest luminosity AGN (10$^{42}$ erg s$^{-1} <$ $L_{\rm X} < 10^{43}$ erg s$^{-1}$) and 35\% of the moderate luminosity AGN (10$^{43}$ erg s$^{-1} <$ $L_{\rm 0.5-10keV} < 10^{44}$ erg s$^{-1}$) meet the $W1-W2$ AGN color criterion, this fraction increases appreciably for the moderately-high (10$^{44}$ erg s$^{-1} <$ $L_{\rm 0.5-10keV} < 10^{45}$ erg s$^{-1}$) and high-luminosity AGN ($L_{\rm 0.5-10keV} > 10^{45}$ erg s$^{-1}$), with 68\% and 71\% of these populations, respectively, meeting the $W1-W2$ AGN definition. In total, 55\% of the Stripe 82X AGN meet the $W1-W2$ AGN color cut. But does this $W1-W2$ color cut select the obscured portion of the X-ray AGN population?

\begin{figure}[ht]
{\includegraphics[scale=0.4,angle=90]{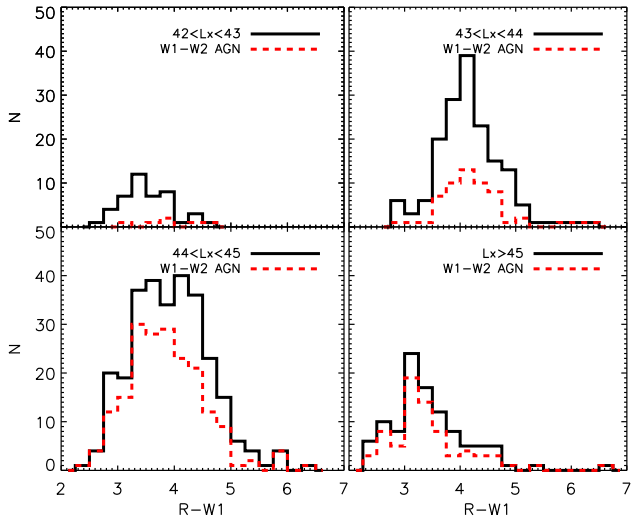}}
\caption[]{\label{w1w2_r-w1} $R-W1$ distributions for different luminosity bins, with the $W1-W2$ identified AGN  highlighted in red. These latter sources span a range of optical-infrared colors and are not preferentially the reddened population. The difference between the red and black histogram at $R-W1>4$ illustrates the reddened AGN missed by the $W1-W2$ selection that can be recovered by $R-W1$.}
\end{figure}

To answer this, we again plot the distribution of $R-W1$ for the different luminosity bins in Figure \ref{w1w2_r-w1} and highlight the subset that are classified as AGN based on their $W1-W2$ colors. Like the parent X-ray AGN samples, $W1-W2$ AGN have a wide range of $R-W1$ values and are not preferentially reddened. While $W1-W2$ is an efficient and robust diagnostic to select unobscured and obscured AGN, including Compton-thick AGN that can be missed altogether by X-ray surveys, additional metrics, such as $R-W1$, are useful in recovering a reddened population missed by this infrared-only cut \citep{stern12,yan2013,donoso}.

%\FloatBarrier

\subsection{\label{hr_r-w1} Is X-ray Obscuration Related to Optical Reddening?}
\begin{figure}[ht]
\begin{center}
{\includegraphics[scale=0.4,angle=90]{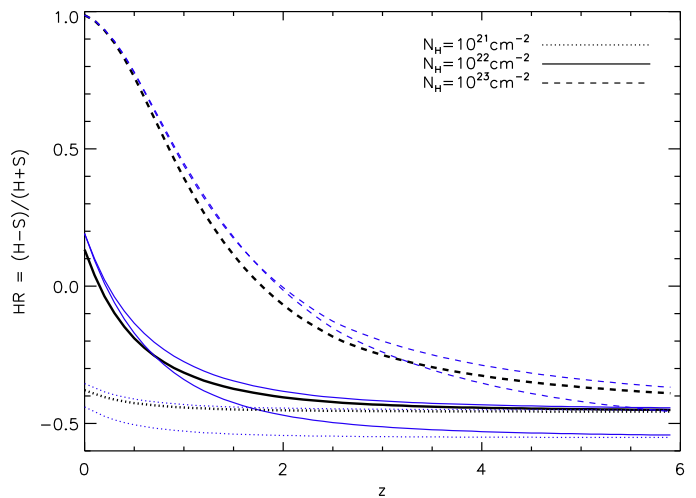}}
\caption[]{\label{hr_v_z} X-ray hardness ratios (HR = $(H-S)/(H+S)$) as a function of redshift for different column densities as indicated in the legend; this is based on an absorbed power law model where $\Gamma$=1.8. The thicker black lines show the relationships for {\it Chandra} while the thinner blue lines correspond to {\it XMM-Newton} MOS (top) and PN (bottom). HR as a proxy of X-ray obscuration is strongly dependent on redshift.}
\end{center}
\end{figure}

A detailed investigation of X-ray obscuration can only be obtained by fitting high-quality X-ray spectra with hundreds to thousands of counts \citep[e.g.,][]{tozzi,akylas,mainieri,lanzuisi,me2014}. As the requisite signal-to-noise for such an analysis is not available for the majority of sources discovered in X-ray surveys, a simpler diagnostic, the hardness ratio (HR), is often used as a proxy of X-ray extinction \citep[e.g.,][]{kim,wilkes,fiore2008,fiore2009,civano12}. HR is defined as $(H - S)/(H + S)$, where $H$ and $S$ are the net counts in the hard (2-10 keV) and soft (0.5-2 keV) bands, respectively.\footnote{The {\it Chandra} hard band is 2-7 keV.} HR as an obscuration diagnostic is redshift-dependent, since higher energy X-ray emission, which is less affected by absorption, is shifted into the observed frame while any soft excess X-ray emission is shifted out of the observed frame as redshift increases. Additionally, the effective area of the instrument plays a role, with different conversions between both {\it Chandra} and {\it XMM-Newton}, as well as between the {\it XMM-Newton} PN camera and {\it XMM-Newton} MOS1 and MOS2 detectors. We illustrate this in Figure \ref{hr_v_z} where we plot HR as a function of redshift for X-ray column density ($N_{H}$) values of $10^{21}$,$10^{22}$, and $10^{23}$ cm$^{-2}$ separately for {\it Chandra} (thick black line) and the {\it XMM-Newton} MOS and PN detectors (thinner blue lines); here we assumed an absorbed powerlaw model where $\Gamma$=1.8. The spectral slope of a source also influences the HR, but obscuration affects the shape of the spectrum, so due to this degeneracy, we do not separately consider the effects of varying the spectral slope. At a fixed redshift, greater HR values correspond to larger column densities \citep[see also][]{weigel}.

To calculate HRs for the Stripe 82X AGN, we use the Bayesian Estimator of Hardness Ratios \citep[BEHR,][]{behr}. This method robustly calculates the HRs and associated errors, even in the low count regime and in cases where a source is undetected in either the hard or soft band. Both the source and background photons are modeled as independent Poisson variables, with posterior distributions obtained by Monte Carlo integration via the Gibbs algorithm. For reference, we have chosen to use the default BEHR values of 10000 draws with the number of burn-in draws set to 5000. 

The number of source and background photons inputted into BEHR for the {\it XMM-Newton} sources were extracted from the coadded PN, MOS1, and MOS2 images. Since the effective exposure times can vary among the detectors, we do not average the theoretical HR values shown in Figure \ref{hr_v_z}, but note that the observed HRs fall within the range depicted. The median HR error bars, $\sim$0.3, span a wider range than the spread in the model {\it XMM-Newton} HRs. 

\begin{figure}[ht]
\begin{center}
{\includegraphics[scale=0.4,angle=90]{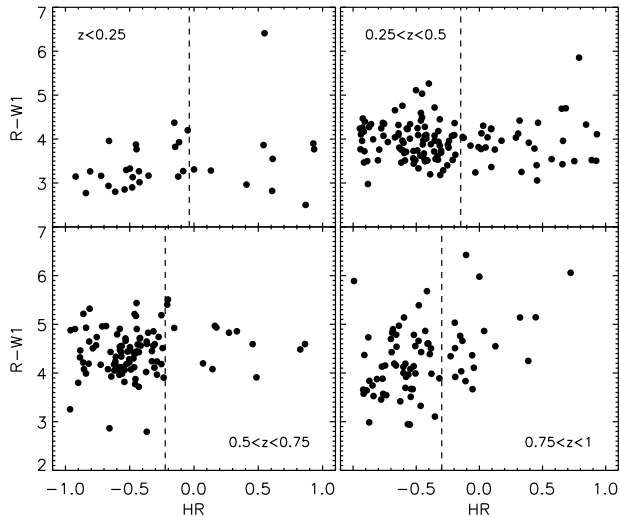}}
\caption[]{\label{hr_v_rw1} $R-W1$ as a function of X-ray hardness ratio for Stripe 82X sources in the redshift ranges ({\it top left}) $z<0.25$, ({\it top right}) $0.25<z<0.5$, ({\it bottom left}) $0.5<z<0.75$, and ({\it bottom right}) $0.75<z<1$. The dashed line denotes the {\it Chandra} HR that corresponds to $N_{H} = 10^{22}$ cm$^{-2}$ for the median redshift of the sub-sample, where HRs greater than this value indicate larger amounts of X-ray obscuration. No trends between optical reddening and X-ray obscuration are apparent. }
\end{center}
\end{figure}

\begin{figure}[ht]
\begin{center}
{\includegraphics[scale=0.4,angle=90]{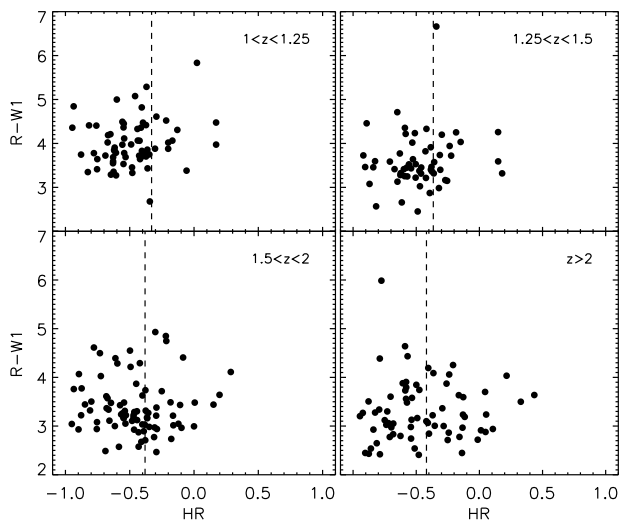}}
\caption[]{\label{hr_v_rw2} Similar to Figure \ref{hr_v_rw1} but for redshift ranges ({\it top left}) $1<z<1.25$, ({\it top right}) $1.25<z<1.5$, ({\it bottom left}) $1.5<z<2$, and ({\it bottom right}) $z>2$. Again, optical obscuration, as parameterized by $R-W1$, is unrelated to X-ray absorption as implied by HR.}
\end{center}
\end{figure}

Since HR as a proxy of X-ray obscuration is heavily dependent on the redshift of the AGN, we investigate the relationship between HR and $R-W1$ in redshift bins. These bins were chosen to be fine enough so that the implied N$_{H}$ value does not vary greatly from the lowest to highest redshift in the bin (see Figure \ref{hr_v_z} for reference), while still being wide enough to include a sufficient number of sources to search for trends. In Figures \ref{hr_v_rw1} - \ref{hr_v_rw2}, we also plot a dashed-line to indicate the {\it Chandra} HR that corresponds to $N_{H}=10^{22}$ cm$^{-2}$ at the median redshift of the bin; HRs greater than this value correspond to higher column densities. As Figures \ref{hr_v_rw1} - \ref{hr_v_rw2} show, no trends exist between optical reddening and X-ray obscuration.

Previous studies of X-ray selected AGN have found that reddened AGN tend to be associated with higher levels of X-ray obscuration, based on their HRs \citep[e.g.,][]{mignoli2004,brusa1,fiore2009,civano12}. Though we do not find a global trend between $R-W1$ and HR, we test whether the redder sources in each redshift bin tend to have higher HRs than the bluer sources. Here, we define reddened AGN as those objects with $R-W1>4$ since unobscured quasars generally do not exceed this value beyond the local Universe (see Figure \ref{seds1}-\ref{r-w1_v_z}). As shown in Table \ref{hr_rw_table}, the reddest AGN do not have higher HRs than the bluer AGN. The apparent dichotomy between our results and those of previous surveys can be largely attributed to the definition of obscured AGN. The extremely reddened AGN studied in \citet{mignoli2004}, \citet{brusa1}, \citet{fiore2009}, and \citet{civano12}, which are associated with larger amounts of X-ray obscuration, are predominantly narrow-lined AGN. However, the Stripe 82X AGN sample consists of mostly broad-lined sources, and are exclusively so at $z>1$. Though some of these broad-lined quasars are reddened, their hardness ratios imply that they are not correspondingly X-ray obscured. Since these AGN are presumably viewed face-on to allow the broad emission lines to be visible, reddening from circumnuclear obscuration would be out of the line-of-sight. The model we have used here to parameterize the AGN spectrum, assuming a foreground screen of extinction, would then be inappropriate. 

Additional complications arise in using HRs as a direct proxy of N$_{H}$. Many X-ray obscured AGN are best described by a partial-covering model, where a fraction of the intrinsic AGN continuum leaks through the circumnuclear obscuration, enhancing emission at soft energies \citep[e.g.,][]{winter,turner,me2009,mayo}. HRs derived via a simple absorbed powerlaw model, where higher values of N$_{H}$ are associated with attenuated soft X-ray emission, would then under-estimate the column density.  Another possibility is that some of these reddened AGN may have ionized, instead of neutral, gas which would also boost the soft X-ray emission \citep[e.g.,][]{turner1991,krolik}. However, these effects decrease with redshift as the rest-frame soft band becomes shifted out of the observed frame, yet we do not see that the reddened AGN in Stripe 82X at higher redshift are associated with higher amounts of implied X-ray obscuration. Finally, we note that while synchrotron emission from jets can cause red $R-W1$ colors \citep[e.g.,][]{massaro}, only 32 of these AGN have a radio counterpart, half of which have bluer colors ($R-W1<4$), suggesting that optical obscuration is responsible for the observed reddening for the majority of the $R-W1>4$ sources.

\begin{table}
\caption{\label{hr_rw_table} Average Hardness Ratios for Reddened and Un-reddened AGN in Stripe 82X}
\begin{tabular}{llll}
\hline
& $\langle$HR$\rangle$ & & $\langle$HR$\rangle$ \\
\hline
         &  $z<0.25$                                 &  & $1<z<1.25$\\
$R-W1>4$ &  0.12 $\pm$ 0.38            &  & -0.46 $\pm$ 0.26 \\
$R-W1<4$ & -0.18 $\pm$ 0.56           &  & -0.51 $\pm$ 0.21 \\
\\
         & $0.25<z<0.5$                          &  & $1.25<z<1.5$   \\
$R-W1>4$ & -0.32 $\pm$ 0.49           &  & -0.42 $\pm$ 0.27 \\
$R-W1<4$ & -0.27 $\pm$ 0.46           &  & -0.49 $\pm$ 0.23 \\
\\
         &   $0.5<z<0.75$                        &  & $1.5<z<2$      \\
$R-W1>4$ & -0.44 $\pm$ 0.35           &  & -0.45 $\pm$ 0.32 \\
$R-W1<4$ & -0.50 $\pm$ 0.34           &  & -0.46 $\pm$ 0.25 \\
\\
         &     $0.75<z<1$ &  & $z>2$          \\
$R-W1>4$ & -0.38 $\pm$ 0.38           &  & -0.41 $\pm$ 0.32 \\
$R-W1<4$ & -0.60 $\pm$ 0.24           &  & -0.46 $\pm$ 0.33 \\

\hline

\end{tabular}
\end{table}

%\FloatBarrier

\section{Conclusions}
Using the 661 {\it Chandra} and {\it XMM-Newton} sources from the initial $\sim$16.5 deg$^2$ release of the Stripe 82X survey that have SDSS counterparts, significant {\it WISE} W1 detections (SNR $\geq5$), and spectroscopic redshifts \citep{me1,me2}, we investigated $R-W1$ as a diagnostic to uncover obscured AGN candidates in wide-area X-ray surveys. We focused on observed quantities to explore the parameter space in which different classes of objects live. $R-W1$ is well correlated with $R-K$, an oft-used reddened AGN selection criterion \citep[e.g.,][]{brusa1,glikman2,glikman3,glikman4,georgakakis,banerji1,banerji2}, and is thus helpful in such searches where $K$-band coverage deeper than 2MASS is lacking. Though $R-W1$ identifies AGN similar to those selected based on $R-K$ and $R-[3.6]\mu m$ colors, the full-sky coverage of {\it WISE}, 35\% sky coverage of SDSS, and 50\% sky coverage of Pan-STARRs, make $R-W1$ a particularly suitable diagnostic to identify interesting AGN candidates for spectroscopic follow-up over a much larger area of the sky. The pre-existing availability of $R-W1$ is of particular relevance for current wide-area X-ray surveys as well as the upcoming all-sky {\it eROSITA} X-ray survey, and will thus recover AGN missed via alternative selection criteria, helping to complete the census of supermassive black hole growth.

Our main results pertaining to the utility of $R-W1$ as an AGN color selection, and the relationships between $R-W1$ and additional multi-wavelength information, are summarized below:

\begin{itemize}

\item Most stars occupy a distinct track in $R-K$ vs. $R-W1$ parameter space, following the relation $R-W1 = 0.998 (\pm0.02) \times (R-K) + 0.18$ (Figure \ref{r-w1_v_r-k_a}, top right). A combination of optical, mid-infrared, and near-infrared colors can then be used to identify stars with high confidence in X-ray samples in the absence of spectroscopic data or to more precisely target extragalactic (or Galactic) sources when undertaking follow-up spectroscopic campaigns. This selection criterion will lead to cleaner samples than employing a simple $R-K$ color cut, and is useful when $W3$-band detections are lacking, making the {\it WISE} $W1-W2$ versus $W2-W3$ color diagnostic \citep{wise} untenable.

\item At redshifts above 0.5, optically normal galaxies hosting AGN (objects lacking broad spectral lines but having X-ray luminosities exceeding 10$^{42}$ erg s$^{-1}$) typically have red $R-W1$ colors (Figure \ref{r-w1_v_r-k_a}, bottom right). Most of these moderate to moderately high luminosity AGN ($10^{43}$ erg s$^{-1} <L_{\rm 0.5-10keV} < 10^{45}$ erg s$^{-1}$) are optically elusive AGN (having narrow emission lines consistent with star-forming galaxies or LINERS) or XBONGs (with optical spectra of passive galaxies, lacking emission lines). We suggest that a color cut of $R-W1>4$ in X-ray samples from shallow surveys separates these elusive AGN at redshifts $0.5 < z < 1$ from galaxies hosting dormant black holes at $z<0.5$. Based on the Stripe 82X sample, the stellar contamination above $R-W1>4$ is $\sim$3\%, but drops to under 0.9\% when using $R-W1>4$ in tandem with $R-K$ and the relationship derived above to remove objects in the stellar locus.

\item We find that the $W1-W2$ color cut to select AGN from \citet{stern12} and \citet{assef} more efficiently identifies the high-luminosity AGN ($L_{\rm 0.5-10keV} > 10^{44}$ erg s$^{-1}$), while most of the AGN below these luminosities are not within the $W1-W2$ AGN locus (Figure \ref{w1w2_lum}). Additionally, $W1-W2$ selected AGN have a range of $R-W1$ colors (Figure \ref{w1w2_lum}), demonstrating that this diagnostic selects both unobscured and obscured AGN; $R-W1$ can then recover portions of the reddened AGN population not selected via $W1-W2$. Optically elusive AGN generally are not within the $W1-W2$ AGN locus.

\item No clear trend exists between implied X-ray obscuration (parameterized by the X-ray hardness ratio) and $R-W1$ (Figures \ref{hr_v_rw1}-\ref{hr_v_rw2}). This lack of correlation likely results from the fact that many of the Stripe 82X AGN are broad-lined objects, where circumnuclear obscuration affecting X-ray emission is presumably occurring out of the line-of-sight and HRs as a proxy for $N_{H}$ is of limited utility when the X-ray spectra are complex. Though our results are affected by incompleteness, it seems clear that an AGN can be dust reddened without being correspondingly X-ray obscured along the line-of-sight.

\end{itemize}

Additionally, we explored the $R-W1$ colors for different observed X-ray luminosity bins (Figure \ref{r-w1_hist} - \ref{r-w1_hist_fit}). Though these trends may change when a higher percentage of X-ray sources are identified with spectroscopic and photometric redshifts, these findings are likely to represent the AGN demographics of sources initially identified in future wide-area X-ray surveys, including {\it eROSITA}, since these samples will have similar, or lower, spectroscopic completeness:

\begin{itemize}

\item The lowest and highest X-ray luminosity bins ($10^{42}$ erg s$^{-1} <$ $L_{\rm 0.5-10keV} < 10^{43}$ erg s$^{-1}$, $L_{\rm 0.5-10eV} > 10^{45}$ erg s$^{-1}$) predominantly have bluer $R-W1$ colors. While the lowest luminosity AGN are in the local Universe, where $R-W1$ is not an indicator of obscuration, the blue $R-W1$ colors of the highest luminosity AGN in the sample cleanly demonstrate that this population is unobscured.

\item Moderate luminosity AGN ($10^{43}$ erg s$^{-1} <$ $L_{\rm 0.5-10keV} < 10^{44}$ erg s$^{-1}$) tend to have redder $R-W1$ colors while moderately high luminosity AGN ($10^{44}$ erg s$^{-1} <$ $L_{\rm 0.5-10keV} < 10^{45}$ erg s$^{-1}$) span the widest range of $R-W1$ colors. In this latter luminosity bin, AGN identified by our ground-based campaign largely represent the red (i.e., obscured) portion of the population. 
\end{itemize}

These trends are only tested on the Stripe 82X sources that we are able to identify via spectroscopic redshifts, which is currently $\sim$59\% of the X-ray sources with SDSS counterparts within the magnitude limits of $r < 22.2$ and $i < 21.3$ and $W1$ band detections at the 5$\sigma$ level. There may exist a population of heavily reddened, luminous AGN in our sample that may be revealed after additional spectroscopic follow-up. Indeed, the most obscured population may not be detected in optical surveys and would by definition be omitted from this sample. We are executing a ground-based near-infrared campaign to target this class of objects, and have discovered several $L_{\rm 0.5-10keV} > 10^{44} - 10^{45}$ erg s$^{-1}$ quasars that are heavily reddened (LaMassa et al. {\it in prep.}). Future spectroscopic follow-up will reveal whether such high luminosity obscured black hole growth is significant.

Finally, we re-emphasize that the relationships and trends reported here pertain to X-ray sources. For instance, non-X-ray emitting stars could populate a much larger range in the $R-K$ versus $R-W1$ parameter space. Dusty star-forming galaxies and (ultra) luminous infrared galaxies have red $R-W1$ colors, but do not necessarily host accreting supermassive black holes. Additionally, as Stripe 82X has relatively high X-ray flux limits, the normal galaxies detected are in the nearby Universe: deep surveys, like the {\it Chandra} Deep Field South \citep{giacconi,xue} uncover normal galaxies much further away \citep{lehmer_gal} which could potentially have colors as red as the Stripe 82X XBONGs and optically elusive galaxies between $0.5 < z < 1$. The results here are instructive in informing follow-up for the many current and planned wide-area, shallow X-ray surveys by developing focused selection criteria for specific classes of interesting objects (e.g., X-ray emitting stars, XBONGs, red quasars) that can be discovered via the power of combining X-ray, infrared, and optical coverage.

\acknowledgments 
SML acknowledges support from grant number NNX15AJ40G. MB acknowledges support from the FP7 Career Integration Grant ``eEASy'' (CIG 321913). Support for TM comes from a National Research Council Research Associateship Award at the Naval Research Laboratory. KS gratefully acknowledges support from Swiss National Science Foundation Professorship grant PP00P2\_138979/1. Support for the work of E. T. was provided by the Center of Excellence in Astrophysics and Associated Technologies (PFB 06) and by the CONICYT Anillo project ACT1101. 

Funding for the SDSS and SDSS-II has been provided by the Alfred P. Sloan Foundation, the Participating Institutions, the National Science Foundation, the U.S. Department of Energy, the National Aeronautics and Space Administration, the Japanese Monbukagakusho, the Max Planck Society, and the Higher Education Funding Council for England. The SDSS Web Site is http://www.sdss.org/.

The SDSS is managed by the Astrophysical Research Consortium for the Participating Institutions. The Participating Institutions are the American Museum of Natural History, Astrophysical Institute Potsdam, University of Basel, University of Cambridge, Case Western Reserve University, University of Chicago, Drexel University, Fermilab, the Institute for Advanced Study, the Japan Participation Group, Johns Hopkins University, the Joint Institute for Nuclear Astrophysics, the Kavli Institute for Particle Astrophysics and Cosmology, the Korean Scientist Group, the Chinese Academy of Sciences (LAMOST), Los Alamos National Laboratory, the Max-Planck-Institute for Astronomy (MPIA), the Max-Planck-Institute for Astrophysics (MPA), New Mexico State University, Ohio State University, University of Pittsburgh, University of Portsmouth, Princeton University, the United States Naval Observatory, and the University of Washington.

Funding for SDSS-III has been provided by the Alfred P. Sloan Foundation, the Participating Institutions, the National Science Foundation, and the U.S. Department of Energy Office of Science. The SDSS-III web site is http://www.sdss3.org/.

SDSS-III is managed by the Astrophysical Research Consortium for the Participating Institutions of the SDSS-III Collaboration including the University of Arizona, the Brazilian Participation Group, Brookhaven National Laboratory, Carnegie Mellon University, University of Florida, the French Participation Group, the German Participation Group, Harvard University, the Instituto de Astrofisica de Canarias, the Michigan State/Notre Dame/JINA Participation Group, Johns Hopkins University, Lawrence Berkeley National Laboratory, Max Planck Institute for Astrophysics, Max Planck Institute for Extraterrestrial Physics, New Mexico State University, New York University, Ohio State University, Pennsylvania State University, University of Portsmouth, Princeton University, the Spanish Participation Group, University of Tokyo, University of Utah, Vanderbilt University, University of Virginia, University of Washington, and Yale University.

This publication makes use of data products from the Wide-field Infrared Survey Explorer, which is a joint project of the University of California, Los Angeles, and the Jet Propulsion Laboratory/California Institute of Technology, funded by the National Aeronautics and Space Administration.

\end{document}